\documentclass[aps,prb,twocolumn,showpacs,groupedaddress]{revtex4}

\usepackage{graphicx}
\usepackage{latexsym}
\usepackage{amssymb}

\begin{document}


\title{Aging dynamics in reentrant ferromagnet: Cu$_{0.2}$Co$_{0.8}$Cl$_{2}$-FeCl$_{3}$
graphite bi-intercalation compound}

\author{Masatsugu Suzuki }
\email[]{suzuki@binghamton.edu}
\affiliation{ Department of Physics, State University of New York
at Binghamton, Binghamton, New York 13902-6000}

\author{Itsuko S. Suzuki }
\affiliation{ Department of Physics, State University of New York
at Binghamton, Binghamton, New York 13902-6000}


\date{\today}

\begin{abstract}
Aging dynamics of a reentrant ferromagnet
Cu$_{0.2}$Co$_{0.8}$Cl$_{2}$-FeCl$_{3}$ graphite bi-intercalation compound
has been studied using AC and DC magnetic susceptibility.  This compound
undergoes successive transitions at the transition temperatures $T_{c}$
($= 9.7$ K) and $T_{RSG}$ ($= 3.5$ K).  The relaxation rate
$S(t)$ exhibits a characteristic peak at $t_{cr}$ close to a wait time
$t_{w}$ below $T_{c}$, indicating that the aging phenomena occur in both
the reentrant spin glass (RSG) phase below $T_{RSG}$ and the ferromagnetic
(FM) phase between $T_{RSG}$ and $T_{c}$.  The relaxation rate $S(t)$
($=\text{d}\chi_{ZFC}(t)/\text{d}\ln t$) in the FM phase exhibits two peaks
around $t_{w}$ and a time much shorter than $t_{w}$ under the positive
$T$-shift aging, indicating a partial rejuvenation of domains.  The aging
state in the FM phase is fragile against a weak magnetic-field
perturbation.  The time ($t$) dependence of $\chi_{ZFC}(t)$ around $t
\approx t_{cr}$ is well approximated by a stretched exponential relaxation:
$\chi_{ZFC}(t) \approx \exp[-(t/\tau)^{1-n}]$.  The exponent $n$ depends on
$t_{w}$, $T$, and $H$.  The relaxation time $\tau$ ($\approx t_{cr}$)
exhibits a local maximum around 5 K, reflecting a chaotic nature of the FM
phase.  It drastically increases with decreasing temperature below
$T_{RSG}$.  
\end{abstract}

\pacs{75.50.Lk, 75.40.Gb, 75.70.Cn, 75.10.Nr}

\maketitle

\section{\label{intro}Introduction}
In random spin systems with competing ferromagnetic and antiferromagnetic
interactions, the spin frustration effect occurs, leading to a spin-glass
(SG) phase at low temperatures.  This situation may change when there is a
majority of the ferromagnetic interactions and a minority of the
antiferromagnetic interactions to create substantial spin frustration
effect.  The system (so-called reentrant ferromagnet) exhibits two phase
transitions at $T_{RSG}$ and $T_{c}$ ($T_{c}>T_{RSG}$): the reentrant spin
glass (RSG) phase below $T_{RSG}$ and the ferromagnetic (FM) phase between
$T_{RSG}$ and $T_{c}$.  Experimental studies on the dynamic magnetic
properties have been carried out for reentrant ferromagnets such as
(Fe$_{0.20}$Ni$_{0.80}$)$_{75}$P$_{16}$B$_{6}$Al$_{3}$,
\cite{Jonason1996a,Jonason1996b,Jonason1998,Jonason1999}
Cr$_{78}$Fe$_{22}$,\cite{Mitchler1993}
(Fe$_{0.65}$Ni$_{0.35}$)$_{0.882}$Mn$_{0.118}$,\cite{Li1994}
CdCr$_{2x}$In$_{2(1-x)}$S$_{4}$ ($x$ = 0.90, 0.95, and
1.00),\cite{Vincent2000a,Vincent2000b,Dupuis2002} and
Fe$_{0.7}$Al$_{0.3}$.\cite{Motoya2003} For
Cr$_{78}$Fe$_{22}$,\cite{Mitchler1993}
(Fe$_{0.65}$Ni$_{0.35}$)$_{0.882}$Mn$_{0.118}$,\cite{Li1994} and
Fe$_{0.7}$Al$_{0.3}$,\cite{Motoya2003} the RSG phase exhibits aging
phenomena, which are very similar to those observed in the SG phase of
SG systems.  No aging phenomenon has been observed in the FM phase.  For
(Fe$_{0.20}$Ni$_{0.80}$)$_{75}$P$_{16}$B$_{6}$Al$_{3}$,
\cite{Jonason1996a,Jonason1996b,Jonason1998,Jonason1999} in contrast, not
only the RSG phase but also the FM phase exhibit aging phenomena.  A
strikingly increased fragility to the magnitude of the magnetic field is
observed when passing from the low-temperature RSG region into the FM
phase, implying a chaotic nature of the FM phase.  The effect of the
probing field $H$ on the relaxation rate has to be carefully considered. 
The dramatic decrease of the limiting field around $T_{RSG}$ may explain
why other experiments on reentrant ferromagnets have not resolved an aging
behavior in the FM phase.  For CdCr$_{2x}$In$_{2(1-x)}$S$_{4}$ with $x$ =
0.90, 0.95, and 1.0,\cite{Vincent2000a,Vincent2000b,Dupuis2002} the aging
behavior of the low frequency AC susceptibility (absorption 
$\chi^{\prime\prime}$) is observed both in the FM and RSG phases.

Experimental studies on the aging dynamics have been limited to a
macroscopic measurement such as the time evolution of zero-field cooled
susceptibility, thermoremnant susceptibility, and the absorption of the AC
magnetic susceptibility.  Recently Motoya et al.\cite{Motoya2003} have
studied time-resolved small angle neutron scattering on
Fe$_{0.70}$Al$_{0.30}$ in order to probe the microscopic mechanism of slow
dynamics.  The Lorentzian form of the scattering pattern and the
temperature variation of the inverse correlation length below $T_{RSG}$
show that the system is composed of only finite-sized clusters.  The size
of the clusters gradually decreases with decreasing temperature.

Cu$_{0.2}$Co$_{0.8}$Cl$_{2}$-FeCl$_{3}$ graphite bi-intercalation compound
(GBIC) is one of typical 3D Ising reentrant ferromagnets.  It has a unique
layered structure where the Cu$_{0.2}$Co$_{0.8}$Cl$_{2}$ intercalate layer
(= $I_{1}$) and FeCl$_{3}$ intercalate layers (=$I_{2}$) alternate with a
single graphite layer ($G$), forming a stacking sequence
(-$G$-$I_{1}$-$G$-$I_{2}$-$G$-$I_{1}$-$G$-$I_{2}$-$G$-$\cdots$) along the
$c$ axis.  In the Cu$_{0.2}$Co$_{0.8}$Cl$_{2}$ intercalate layer, two kinds
of magnetic ions (Cu$^{2+}$ and Co$^{2+}$) are randomly distributed on the
triangular lattice.  The static and dynamic magnetic properties have been
reported in a previous paper.\cite{Suzuki2004} This compound undergoes
successive transitions at the transition temperatures $T_{c}$ (= 9.7 K) and
$T_{RSG}$ (= 3.5 K).  A prominent nonlinear susceptibility is observed
between $T_{RSG}$ and $T_{c}$, suggesting a chaotic nature of the FM phase.

In this paper we report our experimental study on the aging dynamics of the
RSG and FM phases of Cu$_{0.2}$Co$_{0.8}$Cl$_{2}$-FeCl$_{3}$ GBIC
using DC and AC magnetic susceptibility measurements.  Our system is cooled
from 50 K to $T$ ($<T_{c}$) in the absence of an external magnetic field. 
This ZFC aging protocol process is completed at $t_{a} = 0$, where $t_{a}$
is defined as an age (the total time after the ZFC aging protocol process). 
Then the system is aged at $T$ under $H = 0$ until $t_{a} = t_{w}$, where
$t_{w}$ is a wait time.  The aging behavior of the ZFC magnetic
susceptibility $\chi_{ZFC}(t)$ has been measured under the various aging
processes: (i) a wait time $t_{w}$ ($2.0 \times 10^{3} \leq t \leq 3.0
\times 10^{4}$ sec), $T$ ($1.9 \leq T \leq 9$ K), and $H$ ($1 \leq H \leq
60$ Oe) as parameters, and (ii) the $T$-shift and $H$-shift perturbations. 
The relaxation rate defined by $S(t) = \text{d}\chi_{ZFC}/\text{d}\ln t$
(see Sec.~\ref{backA} for the definition) exhibits a peak at a
characteristic time $t_{cr}$ close to $t_{w}$ below $T_{c}$, indicating the
occurrence of the aging phenomena both in the RSG and FM phases. 
We will also show that the $t$ dependence of $\chi_{ZFC}(t)$ around $t
\approx t_{w}$ is well described by a stretched exponential relaxation
[$\chi_{ZFC}(t) \approx \exp[-(t/\tau)^{1-n}$] (see Sec.~\ref{backB}),
where $n$ is an exponent and $\tau$ is a relaxation time nearly equal to
$t_{cr}$.  We will show that $n$, $\tau$, and $t_{cr}$ depend on $T$,
$t_{w}$, and $H$.  The local maximum of $\tau$ and $t_{cr}$ around 5 K is
observed, reflecting the chaotic nature of the FM phase.  A partial
rejuvenation of the system occurs in $S(t)$ under the positive shift in the
FM phase.

\section{\label{back}Background}
\subsection{\label{backA}Scaling form of $\chi_{ZFC}(t_{w};t+t_{w})$ and
$\chi^{\prime\prime}(\omega,t)$ in the SG phase}
After the SG system is cooled to $T$ ($<T_{SG}$) through the ZFC
aging protocol at $t_{a} = 0$, the size of domain defined by $R_{T}(t_{a})$
grows with the age of $t_{a}$ and reaches $R_{T}(t_{w})$ just before the
field is turned on at $t = 0$ or $t_{a} = t_{w}$.\cite{Lundgren1990} The
aging behavior in $\chi_{ZFC}$ is observed as a function of the observation
time $t$.  After $t = 0$, a probing length $L_{T}(t,t_{w})$ corresponding
to the maximum size of excitation grows with $t$, in a similar way as
$R_{T}(t_{a})$.  The quasi-equilibrium relaxation occurs first through
local spin arrangements in length $L_{T}(t,t_{w}) \ll R_{T}(t_{w})$,
followed by non-equilibrium relaxation due to domain growth, when
$L_{T}(t,t_{w}) \approx R_{T}(t_{w})$, so that a crossover between the
short-time quasi-equilibrium decay and the non-equilibrium decay at longer
observation times is expected to occur near $t \approx t_{w}$.  The droplet
model\cite{Fisher1988} predicts the following two.  (i) The equilibrium SG
states at two temperatures with the difference $\Delta T$ are uncorrelated
when the overlap length $L_{\Delta T}$ is smaller than $R_{T}(t_{w})$,
i.e., so-called temperature ($T$)-chaos nature of the SG phase.  (ii) The
equilibrium SG states at two fields with the difference $\Delta H$ are
uncorrelated when the overlap length $L_{\Delta H}$ is smaller than
$R_{T}(t_{w})$.

The absorption $\chi^{\prime\prime}$ of the AC magnetic susceptibility is
evaluated from the spin auto-correlation function $C(t_{a}-t;t_{a}) =
\overline{\langle S_{i} (t_{a}-t)S_{i}(t_{a})\rangle}$ using the
fluctuation-dissipation theorem (FDT)
as\cite{Vincent1997,Komori1999,Picco2001,Berthier2002}
\begin{equation}
\chi^{\prime\prime}(\Delta t_{\omega};t+\Delta t_{\omega})
\approx (\textit{-}\pi/2T)\partial C(\Delta t_{\omega};t+\Delta t_{\omega})
/\partial \ln t,
    \label{eq01}
\end{equation}
where $t_{a}=t+\Delta t_{\omega}$, $\Delta t_{\omega}=2\pi/\omega$
(typically $\Delta t_{\omega} \leq 10^{2}$ sec), $\omega$ is the angular
frequency, and $t$ is much larger than $\Delta t_{\omega}$.  In the
auto-correlation function, the over-line denotes the average over sites and
over different realizations of bond disorder, and the bracket denotes the
average over thermal noises.  For slow processes, the dispersion
$\chi^{\prime}(\Delta t_{\omega};t+\Delta t_{\omega})$ is approximated by
\begin{equation}
\chi^{\prime}(\Delta t_{\omega};t+\Delta t_{\omega})
\approx [1-C(\Delta t_{\omega};t+\Delta t_{\omega})]/T.
    \label{eq02}
\end{equation}
In the quasi-equilibrium regime where the FDT holds, the ZFC susceptibility
$\chi_{ZFC}(t_{w};t+t_{w})$ is described by
\begin{equation}
\chi_{ZFC}(t_{w};t+t_{w}) \approx [1-C(t_{w};t+t_{w})]/T,
    \label{eq03}
\end{equation}
where $t_{a}=t+t_{w}$ and $t_{w}$ is a wait time.  Then the relaxation rate
$S(t)$ is described by
\begin{eqnarray}
S(t) &=& d\chi_{ZFC}(t_{w};t+t_{w})d\ln t \nonumber\\
&=& (-1/T)\partial C(t_{w};t+t_{w})/\partial \ln t,
    \label{eq04}
\end{eqnarray}
which corresponds to $(2/\pi)\chi^{\prime\prime}(t_{w};t+t_{w})$.  In spin
glasses the aging manifests by the fact that $C(t_{w};t_{w}+t)$ shows a
strong dependence on the value of $t_{w}$.

It is predicted that $C(t_{w};t+t_{w})$ can be described by an addition of
the short-time and long-time contributions\cite{Berthier2002}
\begin{equation}
C(t_{w};t+t_{w}) \approx C_{eq}(t)+C_{ag}(t_{w};t+t_{w}),
    \label{eq05}
\end{equation}
whereas the aging at the SG transition temperature $T_{SG}$ leads to a
multiplicative representation,
\begin{equation}
C(t_{w};t+t_{w})\approx C_{eq}(t)C_{ag}(t_{w};t+t_{w}).
    \label{eq06}
\end{equation}
Here the equilibrium part can be described by power-law form
\begin{equation}
C_{eq}(t)\approx t^{-\alpha},
    \label{eq07}
\end{equation}
with a temperature-dependent exponent $\alpha$ ($\approx 0$).  These two
forms (additive and multiplicative representations) are actually not so
different for short times, since $C_{ag}(t_{w};t+t_{w})$ is approximately
constant for $t \ll t_{w}$, in the regime where $C_{eq}(t)$ varies most.

The ZFC susceptibility is described by either an additive form
\begin{equation}
\chi_{ZFC}(t_{w};t+t_{w}) \approx (1/T)[1-C_{eq}(t)-
C_{ag}(t_{w};t+t_{w})],
    \label{eq08}
\end{equation}
or a multiplicative form
\begin{eqnarray}
\chi_{ZFC}(t) &=& \chi_{ZFC}(t_{w};t+t_{w}) \nonumber\\
&\approx & (1/T)[1- C_{eq}(t)C_{ag}(t_{w};t+t_{w})].
    \label{eq09}
\end{eqnarray}
When the additive form of $C$ is used, the absorption
$\chi^{\prime\prime}(\omega,t)$ can be rewritten as
\begin{equation}
\chi^{\prime\prime}(\omega,t)
= \chi_{eq}^{\prime\prime}(t) +\chi_{ag}^{\prime\prime}(\omega,t),
    \label{eq10}
\end{equation}
where
\begin{equation}
\chi_{ag}^{\prime\prime}(\omega,t)
= (-\pi/2T)\partial C_{ag}(\Delta t_{\omega};t+\Delta 
t_{\omega})/\partial \ln t,
    \label{eq11}
\end{equation}
and
\begin{equation}
\chi_{eq}^{\prime\prime}(t) =
(-\pi/2T)\partial C_{eq}(t)/\partial \ln t\approx
t^{-\alpha}.
    \label{eq12}
\end{equation}

Here we assume that the aging contribution $C_{ag}(t_{w};t+t_{w})$ is
approximated by a scaling function of $t/t_{w}$ as\cite{Berthier2002}
\begin{equation}
C_{ag}(t_{w};t+t_{w})=F(t/t_{w}).
    \label{eq13}
\end{equation}
In Eq.(\ref{eq10}) the first term
$\chi_{eq}^{\prime\prime}(t)$ [$\approx t^{-\alpha}$] is independent of
$\omega$, while the second term $\chi_{ag}^{\prime\prime}(\omega t)$
[$\approx (\omega t)^{-b}$] is a function of $\omega t$ (see
Sec.~\ref{resultF} for detail).  Here $\alpha$ ($\approx 0$) and $b$
($\approx 0.2$) are exponents of quasi-equilibrium part and aging parts. 
In all cases we observe a slow time decay (aging) of
$\chi^{\prime\prime}(\omega,t)$ towards an asymptotic frequency dependent
value $\chi^{\prime\prime}(\omega,T)$ (stationary susceptibility):
\begin{equation}
\chi^{\prime\prime}(\omega,t) \rightarrow 
\chi^{\prime\prime}(\omega,T) \approx \omega^{\alpha},
    \label{eq15}
\end{equation}
in the limit of $\omega t \rightarrow \infty$, where
$\chi_{ag}^{\prime\prime}(\omega t)$ is assumed to be zero.

\subsection{\label{backB}Stretched exponential relaxation in the SG phase}

\begin{figure}
\includegraphics[width=7.5cm]{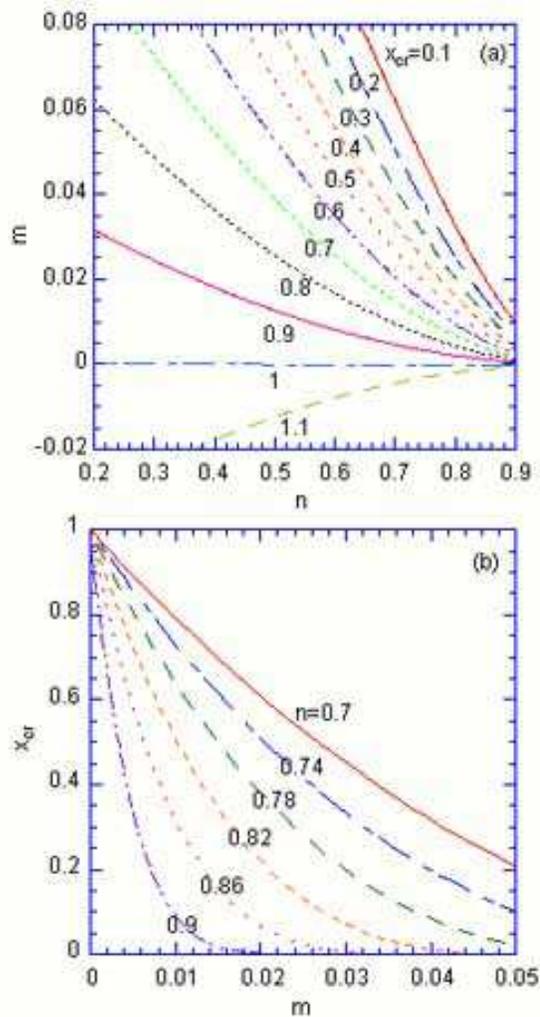}%
\caption{\label{fig01}(Color online)(a) Contour plot of $x_{cr}$ ($0.1 \leq
x_{cr} \leq 1.1$) in the ($n$, $m$) plane, where $x_{cr}=t_{cr}/\tau$, and
the points with the same $x_{cr}$ are connected by the same solid line. 
The definition of $t_{cr}$ and $\tau$ is given in the text.  (b) Plot of
$x_{cr}$ vs $m$ at various $n$.  The expression for $x_{cr}$ is given by
Eqs.(\ref{eq20}) and (\ref{eq21}).}
\end{figure}

Here we present a simple review on the stretched exponential relaxation of
$\chi_{ZFC}$ of SG phase after the ZFC aging
protocol.  Theoretically\cite{Ogielski1985,Koper1988} and experimentally
\cite{Chamberli1984,Hoogerbeets1985a,Hoogerbeets1985b,Alba1986,Lundgren1986,
Alba1987,Granberg1987,Nordblad1987,Hoogerbeets1987,Chu1994,Chu1995} it has
been accepted that the time variation of $\chi_{ZFC}(t)$ may be described
by a product of a power-law form and a stretched exponential function
\begin{equation}
\chi_{ZFC}(t) =M_{ZFC}(t)/H =
\chi_{0}-At^{-m}\exp[-(t/\tau)^{1-n}],
    \label{eq16}
\end{equation}
where the exponent $m$ may be positive and is very close to zero, $n$ is
between 0 and 1, $\tau$ is a characteristic relaxation time, and $\chi_{0}$
and $A$ are constants.  In general, these parameters are dependent on
$t_{w}$.  This form of $\chi_{ZFC}(t)$ incorporates both the nonequilibrium
aging effect through the stretched exponential factor
$[\exp[-(t/\tau)^{1-n}]$ in the crossover region ($t \approx t_{w}$ and
$t>t_{w}$) between the quasi equilibrium state and nonequilibrium state,
and an equilibrium relaxation response at $t \ll t_{w}$ through a pure
power-law relaxation ($t^{-m}$).  Note that
Ogielski\cite{Ogielski1985} fits his data by a stretched exponential multiplied by a
power function.  For $0.6 <T/T_{SG} <
1$, Ogielski\cite{Ogielski1985} fits it by a power law with a different temperature
dependence of exponent $m$.  When $t \ll \tau$, $\chi_{ZFC}(t)$ is well
described by a power law form given by $At^{-m}$.  However, in the regime
of $t \approx \tau$, the stretched exponential relaxation is a very good
approximation in spite of finite $m$ that is very small.

For all temperatures, $\chi_{ZFC}(t)$ increases with increasing $t$ and the
relaxation rate $S(t)$, which is defined by
\begin{equation}
S(t) =\text{d}\chi_{ZFC}(t)/\text{d}\ln t =t
\text{d}\chi_{ZFC}(t)/\text{d}t,
    \label{eq17}
\end{equation}
exhibits a maximum at $t_{cr}$ that is close to $t_{w}$.  Using
Eq.(\ref{eq16}) for $\chi_{ZFC}(t)$, the relaxation rate $S(t)$ can be
derived as
\begin{eqnarray}
S(t) &=& -(A\tau^{-m})\exp[-(t/\tau)^{1-n}]\nonumber\\
& & (t/\tau)^{-(m+n)}[(1-n)+m(t/\tau)^{n-1}].
    \label{eq18}
\end{eqnarray}
The condition that $S(t)$ may have a peak at $t = t_{cr}$
[$\text{d}S(t)/\text{d}t = 0$] leads to the ratio $x_{cr}=t_{cr}/\tau$
satisfying the following equation
\begin{equation}
(1-n)^{2}x_{cr}^{2}-(1-n)(1-2m-n)x_{cr}^{n+1}+m^{2}x_{cr}^{2n}=0.
    \label{eq19}
\end{equation}
The solution of Eq.(\ref{eq19}) can be exactly obtained as
\begin{equation}
x_{cr}=t_{cr}/\tau=(\xi /2)^{1/(1-n)},
    \label{eq20}
\end{equation}
with
\begin{equation}
\xi = [1-2m-n+(1-n)^{1/2}(1-4m-n)^{1/2}]/(1-n),
    \label{eq21}
\end{equation}
where $4m+n<1$.  Note that the value of $x_{cr}$ is uniquely determined
only by the values of $n$ and $m$.  When $m = 0$, $x_{cr} = 1$ (or
$t_{cr}=\tau$), which is independent of $n$.  Figure \ref{fig01}(a) shows
the contour plot of $x_{cr}$ in the ($n$,$m$) plane with $-0.02 \leq m \leq
0.08$ and $0.2 \leq n \leq 0.9$, where the points having the same $x_{cr}$
are connected by each solid line.  The value of $x_{cr}$ is lower than 1
for $m>0$, is equal to 1 for $m = 0$ irrespective of $n$, and is larger
than 1 for $m<0$.  Figure \ref{fig01}(b) shows a plot of $x_{cr}$ as a
function of $m$ at various fixed $n$.  The maximum value of $S(t)$ at $t =
t_{cr}$ is given by
\begin{eqnarray}
S_{\max}& &=A\tau ^{-m} 2^{-1+\frac{m}{1-n}}\nonumber\\
& &\exp \lbrack
-\frac{1}{2} +\frac{m}{1-n} -\frac{\sqrt{1-4m-n}}{2\sqrt{1-n}}
\rbrack (1-n)^{\frac{m}{1-n}}\nonumber\\
& &\times (1-n+\sqrt{1-n} \sqrt{1-4m-n} )\nonumber\\
& &\times (1-2m-n+\sqrt{1-n}\sqrt{1-4m-n} )^{\frac{-m}{1-n} }.
    \label{eq22}
\end{eqnarray}

When $m = 0$, $S_{max}$ is equal to $S_{max}^{0}$ [$=A(1-n)/e$] with $e =
2.7182$.

\section{\label{exp}EXPERIMENTAL PROCEDURE}
We used the same sample of Cu$_{0.2}$Co$_{0.8}$Cl$_{2}$-FeCl$_{3}$ GBIC
that was used in the previous paper.\cite{Suzuki2004} The detail of the sample
characterization and synthesis were presented in the references of the 
previous paper.\cite{Suzuki2004}  The DC magnetization
and AC susceptibility of Cu$_{0.2}$Co$_{0.8}$Cl$_{2}$-FeCl$_{3}$ GBIC were
measured using a SQUID magnetometer (Quantum Design, MPMS XL-5) with an
ultra low field capability option.  The remnant magnetic field was reduced
to zero field (exactly less than 3 mOe) at 298 K for both DC magnetization
and AC susceptibility measurements.  The AC magnetic field used in the
present experiment has a peak magnitude of the AC field ($h$) and a
frequency $f =\omega /2\pi$.  Each experimental procedure for measurements
is presented in the text and figure captions.

In our measurement of the time ($t$) dependence of the zero-field cooled
(ZFC) magnetization ($M_{ZFC}$), the time required for the ZFC aging
protocol and subsequent wait time were precisely controlled.  Typically it
takes $240 \pm 3$ sec to cool the system from 10 K to $T$ ($<T_{c}$).  It
takes another $230 \pm 3$ sec until $T$ becomes stable within the
uncertainty ($\pm 0.01$ K).  The system is kept at $T$ and $H = 0$ for a
wait time $t_{w}$ (typically $t_{w} = 1.5 \times 10^{4}$ or $3.0 \times
10^{4}$ sec).  At time $t = 0$, external magnetic field $H$ is applied
along any direction perpendicular to the $c$ axis.  The time for setting up
a magnetic field from $H$ = 0 to $H = 1$ Oe is $68 \pm 2$ sec.  In the ZFC
measurement, the sample is slowly moved through the pick-up coils over the
scan length (4 cm).  The magnetic moment of the sample induces a magnetic
flux change in the pick-up coils.  It takes 12 sec for each scan.  The data
at $t$ is regarded as the average of $M_{ZFC}$ measured over the scanning
time $t_{s}$ between the times $t-(t_{s}/2)$ and $t-(t_{s}/2)$.  Thus the
time window $\Delta t$ is a scanning time ($t_{s}$).  The measurement was
carried out at every interval of $t_{s}+t_{p}$, where $t_{p}$ is a pause
between consecutive measurements.  Typically we used (i) the time window
$\Delta t = 36$ sec for the measurements with three scans and $t_{p} = 45$ sec or 30 sec, and 
(ii) the time window $\Delta t
= 12$ sec for the measurements with one scan and $t_{p} = 1$ or 2 sec.

\section{\label{result}RESULT}
\subsection{\label{resultA}$\chi^{\prime\prime}(\omega,T)$ and
$\chi_{ZFC}(T)$}

\begin{figure}
\includegraphics[width=7.5cm]{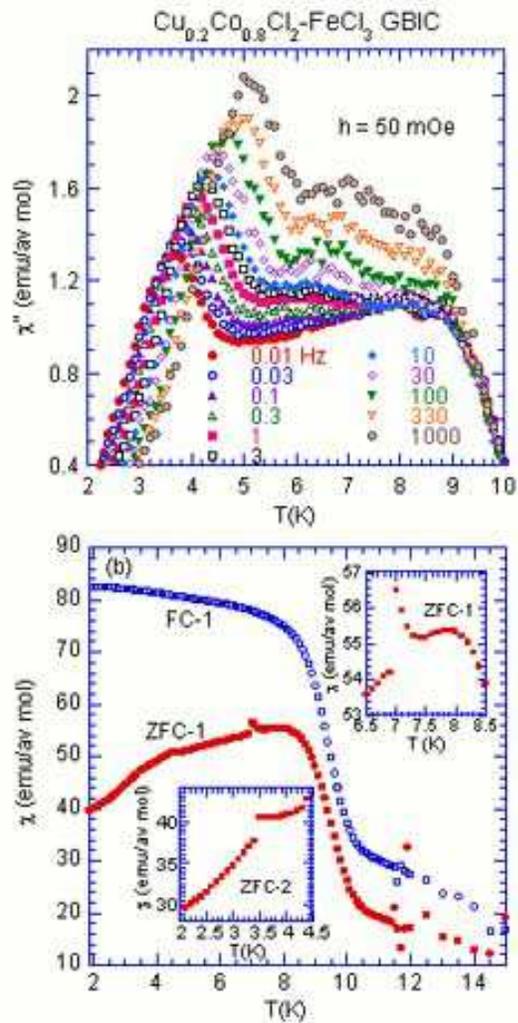}%
\caption{\label{fig02}(Color online)(a) $\chi^{\prime\prime}(\omega,T)$ vs
$T$ at various $f$ ($0.01 \leq f\leq 1000$ Hz).  $h = 50$ mOe.  (b) $T$
dependence of $\chi_{ZFC}$ ({\Large $\bullet$}) and $\chi_{FC}$ ({\Large
$\circ$}) at $H = 1$ Oe for Cu$_{0.2}$Co$_{0.8}$Cl$_{2}$-FeCl$_{3}$ GBIC. The
measurement was carried out after the ZFC aging protocol: annealing of the
system at 50 K for 1200 sec at $H = 0$ and quenching from 50 to 1.9 K.
During the ZFC measurement with increasing $T$ from 1.9 to 20 K, the system
was aging at $T_{1}=7.0$ K (ZFC-1 measurement) and $T_{2}=3.5$ K 
(ZFC-2 measurement) for
wait time $t_{w1} = t_{w2} = 1.5 \times 10^{4}$ sec in the presence of 
$H = 1$ Oe. 
After the ZFC measurement, the system was annealed at $T = 50$ K for $1.2
\times 10^{3}$ sec in the presence of $H$ (= 1 Oe).  The FC measurement
(denoted as FC-1 and FC-2 ) was carried with decreasing $T$ from 20 to 1.9
K. The $T$ dependence of FC-1 susceptibility is the same as that of FC-2
susceptibility.}
\end{figure}

In the previous paper\cite{Suzuki2004} the magnetic properties of
Cu$_{0.2}$Co$_{0.8}$Cl$_{2}$-FeCl$_{3}$ GBIC have been extensively studied
from our results on DC susceptibility ($\chi_{ZFC}$, $\chi_{FC}$), and AC
magnetic susceptibility ($\chi^{\prime}$ and $\chi^{\prime\prime}$).  In
Fig.~\ref{fig02}(a), as a typical example we show the $T$ dependence of the
absorption $\chi^{\prime\prime}(\omega,T)$ at various $f$, where $h = 50$
mOe.  It is concluded from this figure that that our system undergoes two
phase transitions at $T_{RSG} = 3.5$ K and $T_{c} = 9.7$ K. There are a RSG
phase below $T_{RSG}$ and an FM phase between $T_{RSG}$ and $T_{c}$.

In order to demonstrate the evidence of aging behavior in $\chi_{ZFC}$, we
measured the $T$ dependence of $\chi_{ZFC}$ (= $M_{ZFC}/H$) in the
following two ZFC protocols.  The system was quenched from 50 to 1.9 K in
the absence of $H$ before the measurement.  The susceptibility $\chi_{ZFC}$
was measured with increasing $T$ from 1.9 K to $T_{1} = 7.0$ K in the
presence of $H$ (= 1 Oe), where $T_{1}$ is in the FM phase.  The system was
kept at $T_{1}$ for a wait time $t_{w1} = 1.5 \times 10^{4}$ sec. 
Subsequently $\chi_{ZFC}$ was measured with further increasing $T$ from
$T_{1}$ to 20 K. The system was annealed at 50 K for $1.2 \times 10^{3}$ sec in the
presence of $H$ (= 1 Oe).  The susceptibility $\chi_{FC}$ was continuously
measured from 20 to 1.9 K. In Fig.~\ref{fig02}(b) we show the $T$
dependence of $\chi_{ZFC}$ and $\chi_{FC}$ ($H = 1$ Oe) thus obtained
(denoted as the ZFC-1 and FC-1 processes).  We find that there is a
remarkable increase of $\chi_{ZFC}$ with increasing $t$ during the one stop
at $T_{1} = 7.0$ K. As $T$ again increases from $T_{1}$, $\chi_{ZFC}$
starts to decreases with increasing $T$ and merges with $\chi_{ZFC}^{ref}$
without the stop as a reference at $T \approx 7.3$ K (see the inset of
Fig.~\ref{fig02}(b)).  The state at $T = 7.3$ K is uncorrelated with that
at $T_{1}$ since the temperature difference $\Delta T$ is large so that the
domain size $R_{T1}(t_{w1})$ is much larger than the overlap length
$L_{\Delta T}$.

In another measurement (denoted as the ZFC-2 and FC-2 processes), the
system was kept at $T_{2} = 3.5$ K in the RSG phase for $t_{w2}= 1.5 \times
10^{4}$ sec during the ZFC process.  In the inset of Fig.2(b) we show the
$T$ dependence of $\chi_{ZFC}$ in the ZFC-2 process.  There is a remarkable
increase of $\chi_{ZFC}$ during the one stop at $T_{2}$.  As $T$ again
increases from $T_{2}$, $\chi_{ZFC}$ starts to increase with a derivative
d$\chi_{ZFC}$/d$T$ that is positive and much smaller than that of
$\chi_{ZFC}^{ref}$.  This susceptibility $\chi_{ZFC}$ merges with
$\chi_{ZFC}^{ref}$ at $T$ $\approx 3.9$ K. The state at $T = 3.9$ K is
uncorrelated with that at $T_{2}$ since the temperature difference $\Delta
T$ is large so that the domain size $R_{T2}(t_{w2})$ is much larger than
$L_{\Delta T}$.  The $T$ dependence of $\chi_{FC}$ in the FC-2 process is
the same as that in the FC-1 process.  Note that the gradual increase of
$\chi_{ZFC}$ with increasing $T$ from the stop temperature is also observed
in the SG phase of the 3D Ising spin glasses
Fe$_{0.5}$Mn$_{0.5}$TiO$_{3}$\cite{Bernardi2001} and
Cu$_{0.5}$Co$_{0.5}$Cl$_{2}$-FeCl$_{3}$ GBIC.\cite{Suzuki2003} Such an
aging behavior in $\chi_{ZFC}$ may be common to the RSG and SG phases. 
This is in contrast to the gradual decrease of $\chi_{ZFC}$ with increasing
$T$ from the stop temperature in the FM phase.

\subsection{\label{resultB}$T$ dependence of $S(t)$}

\begin{figure}
\includegraphics[width=7.5cm]{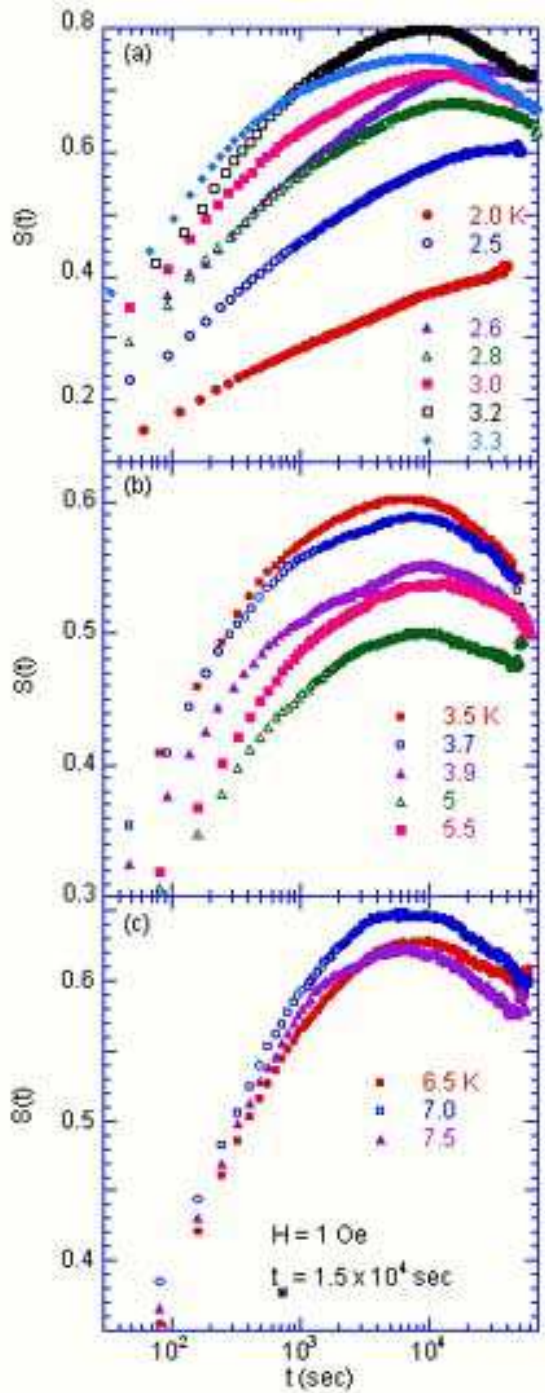}%
\caption{\label{fig03}(Color online)(a) and (b) Relaxation rate $S$
[$=\text{d}\chi_{ZFC}/\text{d}\ln t$] vs $t$ at various $T$.  $H = 1$ Oe. 
$t_{w} = 1.5 \times 10^{4}$ sec.  $H$ is applied along a direction
perpendicular to the $c$ axis.  The ZFC aging protocol: annealing of the
system at 50 K for $1.2 \times 10^{3}$ sec at $H = 0$, quenching from 50 K
to $T$, and then isothermal aging at $T$ and $H = 0$ for a wait time
$t_{w}$.  The measurement was started at $t = 0$ when the field $H$ is
turned on.}
\end{figure}

\begin{figure}
\includegraphics[width=7.5cm]{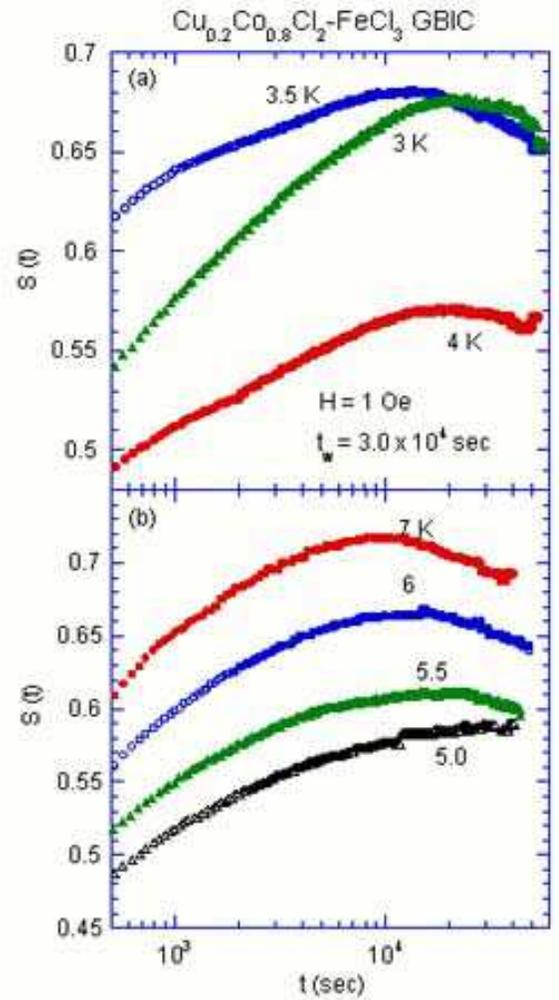}%
\caption{\label{fig04}(Color online)(a) and (b) $S$ vs $t$ at various $T$. 
$H = 1$ Oe.  $t_{w} = 3.0 \times 10^{4}$ sec.}
\end{figure}

\begin{figure}
\includegraphics[width=7.5cm]{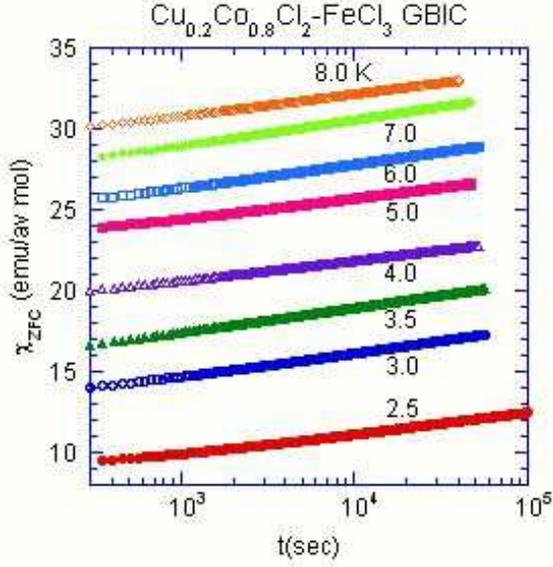}%
\caption{\label{fig05}(Color online)$t$ dependence of $\chi_{ZFC}(t)$ at
various $T$.  $H = 1$ Oe.  $t_{w} = 3.0 \times 10^{4}$ sec.  The solid
curves are fits to Eq.(\ref{eq16}) with $m = 0$ for stretched exponential
relaxation.}
\end{figure}

\begin{figure}
\includegraphics[width=7.5cm]{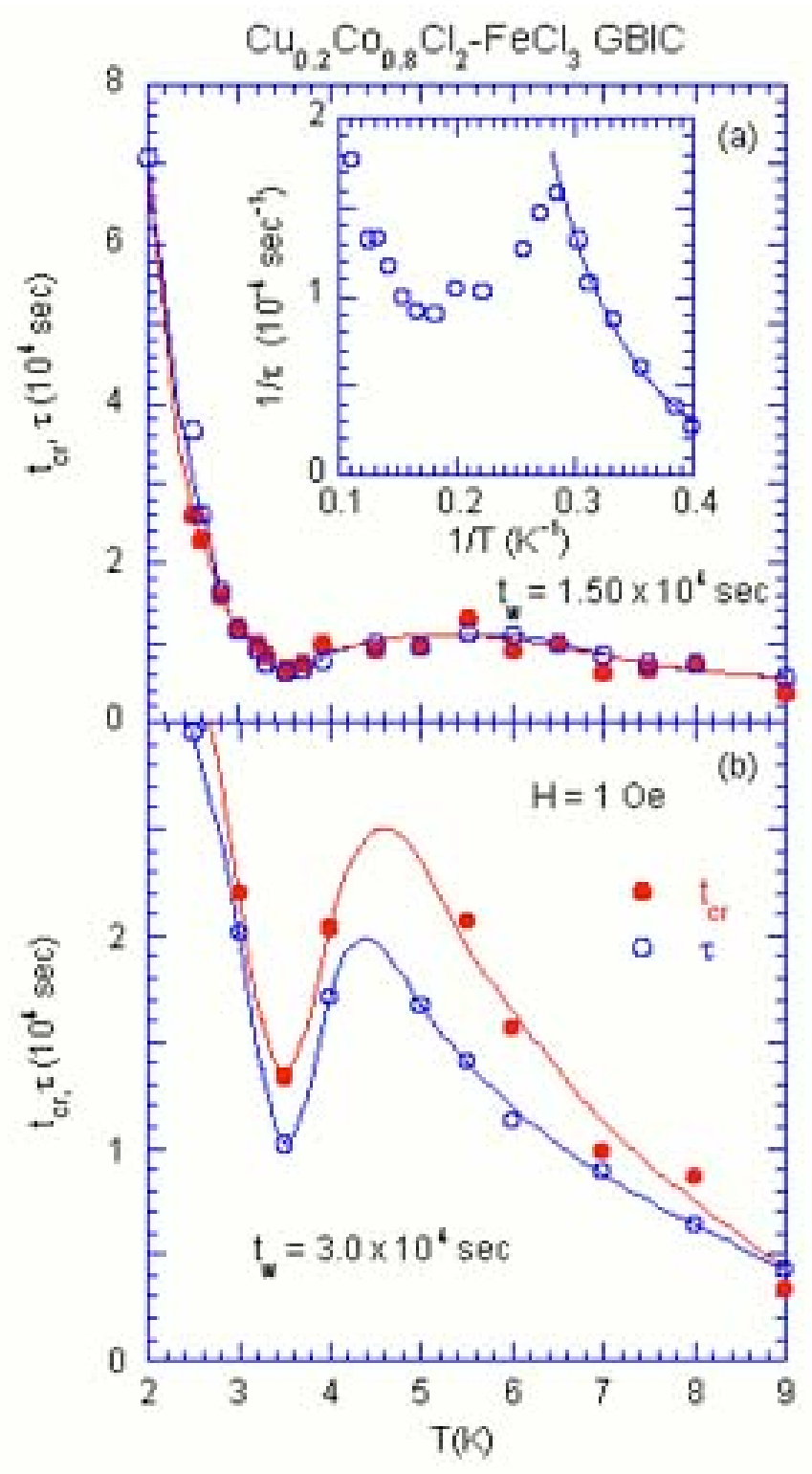}%
\caption{\label{fig06}(Color online)$t_{cr}$ vs $T$ and $\tau$ vs $T$. 
$t_{cr}$ is a characteristic time at which $S(t)$ has a maximum $S_{max}$. 
$\tau$ is the relaxation time for the stretched exponential relaxation.  $H
= 1$ Oe.  (a) $t_{w} = 1.5 \times 10^{4}$ sec.  (b) $t_{w} = 3.0 \times
10^{4}$ sec.  The solid lines are guide to the eyes.  The inset of (a)
shows the plot of $1/\tau$ vs $1/T$ ($T<T_{RSG}$), where the solid line is
a least-squares fit to Eq.(\ref{eq23}).  The fitting parameters are given
in the text.}
\end{figure}

We have measure the $t$ dependence of $\chi_{ZFC}$ under the condition with
various combinations of $T$, $H$, and $t_{w}$.  Figures \ref{fig03} and
\ref{fig04} show the $t$ dependence of the relaxation rate $S(t)$ at
various $T$ ($\leq 2.0 \leq T \leq 9.0$ K) for $t_{w} = 1.5 \times 10^{4}$
sec and $3.0 \times 10^{4}$ sec, respectively, where $H = 1$ Oe.  The
relaxation rate $S(t)$ exhibits a broad peak at a characteristic time
$t_{cr}$ in the FM phase as well as the RSG phase, indicating the aging
behaviors.  In Fig.~\ref{fig05} we show the $t$ dependence of $\chi_{ZFC}$
at $H = 1$ Oe for $t_{w} = 3.0 \times 10^{4}$ sec.  We find that
$\chi_{ZFC}(t)$ is well described by a stretched exponential relaxation
form given by Eq.(\ref{eq16}) with $m = 0$ for $10^{2} \leq t \leq 3.0
\times 10^{4}$ sec.  The least squares fits of these data to
Eq.(\ref{eq16}) with $m = 0$ yields the parameters $\tau$, $n$ and $A$. 
Figure \ref{fig06}(a) shows the $T$ dependence of $t_{cr}$ and $\tau$ for $t_{w} =
1.5 \times 10^{4}$ sec, where $H = 1$ Oe.  The $T$ dependence of $t_{cr}$
almost agrees well with that of $\tau$ for $2 \leq T \leq 9$ K. This
indicates that the relaxation is dominated by a stretched exponential
relaxation with $m = 0$ (see Sec.~\ref{backB}).  The relaxation time $\tau$
($\approx t_{cr}$) shows a broad peak centered around 5.5 K between
$T_{c}$ and $T_{RSG}$, and a local minimum around $T_{RSG}$.  It decreases
with further increasing $T$ below $T_{RSG}$.  As far as we know, there has
been no report on the broad peak of $t_{cr}$ (or $\tau$) in the FM phase of
reentrant ferromagnts.  The existence of the broad peak around 5.5 K
suggests the chaotic nature of the FM phase in our system.  The drastic
increase of $t_{cr}$ (or $\tau$) below $T_{RSG}$ with decreasing $T$ is a
feature common to the SG phases of typical SG systems.  Note that at $T =
2$ K no peak in $S(t)$ is observed for $1 \times 10^{2} \leq t \leq 6 \times 10^{4}$
sec.  The value of $\tau$ is estimated as $7 \times 10^{4}$ sec , which is
much larger than $t_{w}$.  In contrast, as shown in Fig.~\ref{fig06}(b) for
$t_{w} = 3.0 \times 10^{4}$ sec, the value of $t_{cr}$ is larger than that
of $\tau$ in the FM phase.  Both $t_{cr}$ and $\tau$ show a broad peak at
$T$ between 4 and 5 K. This peak temperature is slightly lower than that
for $t_{w} = 1.5 \times 10^{4}$ sec, suggesting that the relaxation
mechanism is dependent on $t_{w}$.  In the inset of Fig.~\ref{fig06}(a) we
show the plot of $1/\tau$ as a function of $1/T$.  In the limited
temperature range ($2.5 \leq T \leq 3.3$ K), $1/\tau$ can be approximated
by an exponential $T$ dependence
\begin{equation}
1/\tau = c_{1}\exp(-c_{2}T_{RSG}/T),
    \label{eq23}
\end{equation}
with $c_{1} =(1.16 \pm 0.32) \times 10^{-2}$ sec$^{-1}$ and $c_{2} = 4.2
\pm 0.2$.  Our value of $c_{2}$ is relatively larger than those derived by
Hoogerbeets et al.  ($c_{2} = 2.5$)\cite{Hoogerbeets1985a,Hoogerbeets1985b}
from an analysis of thermoremanent magnetization (TRM) relaxation
measurements on four SG systems: Ag:Mn (2.6 at.~\%), Ag: Mn (4.1
at.~\%), Ag:[Mn (2.6 at.~\%) + Sb (0.46 at.~\%)], and Cu:Mn (4.0 at.~\%). 
Note that in their work the stretched exponential is taken as
representative of the short time ($t<t_{w}$) relaxation.  One must thus be
careful in comparing the results.

\begin{figure}
\includegraphics[width=7.5cm]{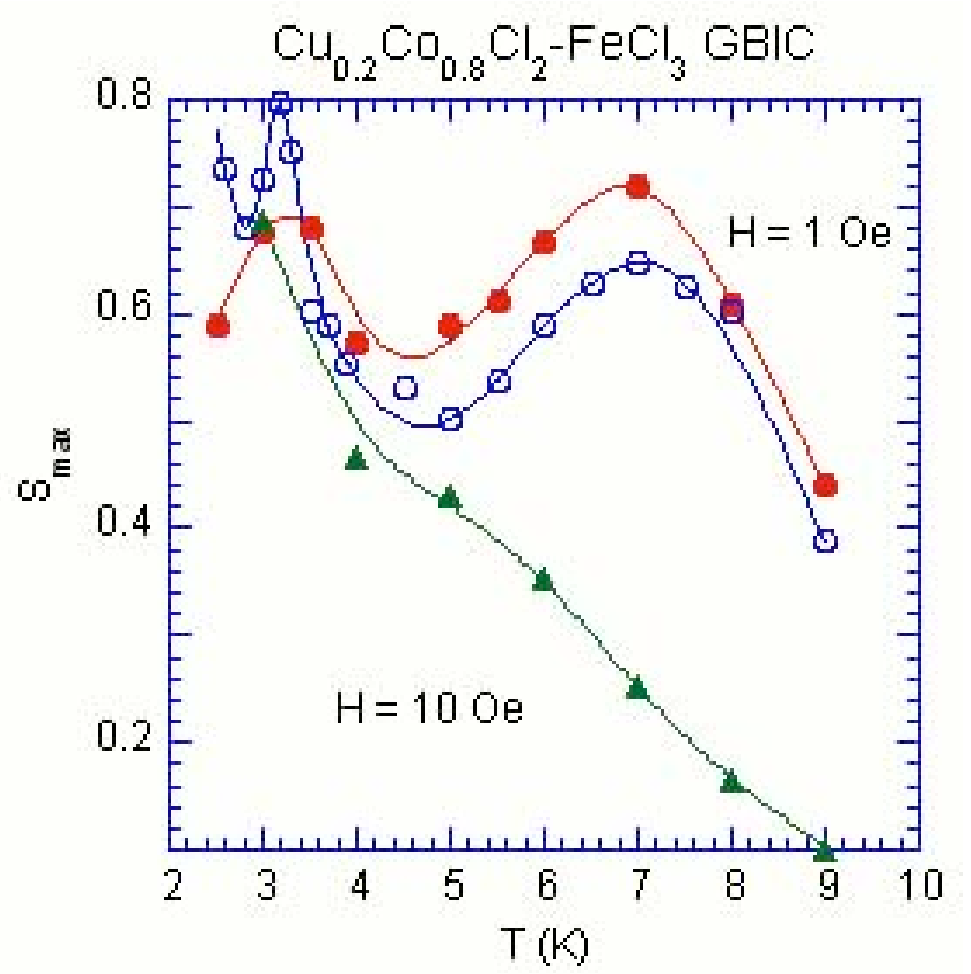}%
\caption{\label{fig07}(Color online)$S_{max}$ vs $T$ for the three cases:
(i) $H = 1$ Oe and $t_{w} = 3.0 \times 10^{4}$ sec ({\Large $\bullet$}),
(ii) $H$ = 1 Oe and $t_{w} = 1.5 \times 10^{4}$ sec ({\Large $\circ$}), and
(iii)) $H$ = 10 Oe and $t_{w} = 2.0 \times 10^{3}$ sec ($\blacktriangle$).  The
soild lines are guides to the eyes.}
\end{figure}

Figure \ref{fig07} shows the $T$ dependence of $S_{max}$ (the peak height
of $S(t)$ at $t = t_{cr}$) for the following three cases: (i) $H = 1$ Oe
and $t_{w} = 3.0 \times 10^{4}$ sec, (ii) $H = 1$ Oe and $t_{w} = 1.5
\times 10^{4}$ sec, and (iii) $H = 10$ Oe and $t_{w} = 2.0 \times 10^{3}$
sec.  The peak height $S_{max}$ at $H = 1$ Oe exhibits two peaks around $T
= T_{RSG}$ and at 7.0 K just below $T_{c}$, independent of $t_{w}$ ($=1.5
\times 10^{4}$ sec or $3.0 \times 10^{4}$ sec).  Note that the peak of
$S_{max}$ at 7.0 K is strongly dependent of $H$: it disappears at $H = 10$
Oe for $t_{w} = 2.0 \times 10^{3}$ sec.

\begin{figure}
\includegraphics[width=7.5cm]{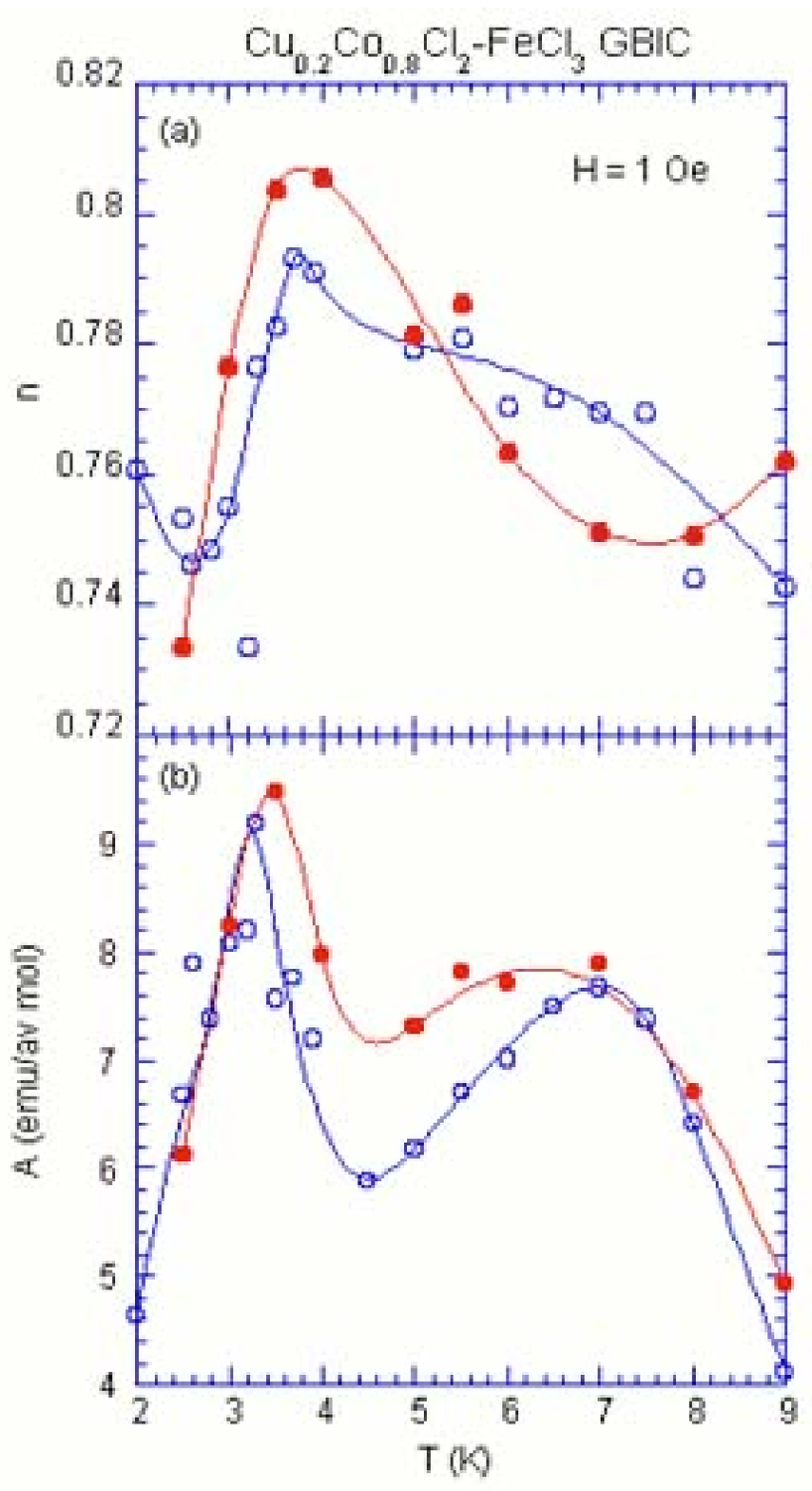}%
\caption{\label{fig08}(Color online)(a) $n$ vs $T$ and (b) $A$ vs $T$,
where $t_{w} = 3.0 \times 10^{4}$ sec ({\Large $\bullet$}) and $1.5 \times
10^{4}$ sec ({\Large $\circ$}).  $H = 1$ Oe.  The soild lines are guides to
the eyes.}
\end{figure}

In Fig.~\ref{fig08}(a) we show the plot of the exponent $n$ as a function
of $T$ for $t_{w} = 1.5 \times 10^{4}$ and $3.0 \times 10^{4}$ sec, where
$H = 1$ Oe.  For $t_{w} = 3.0 \times 10^{4}$ sec the exponent $n$ increases
with increasing $T$ and exhibits a peak at $T = 4$ K just above $T_{RSG}$. 
The exponent $n$ decreases with further increasing $T$, showing a local
minimum around 7.5 K. For $t_{w} = 1.5 \times 10^{4}$ sec, the exponent $n$
exhibits a local minimum ($n \approx 0.75$) around 2.5 K ($T/T_{RSG} =
2.5/3.5 = 0.71$) and a local maximum ($n \approx 0.79$) around 4.0 K. The
local minimum of $n$ below $T_{RSG}$ is a feature common to that of SG
systems, where $n$ has a local minimum at $T = T^{*}$ given by
$T^{*}/T_{SG} \approx 0.70$.\cite{Hoogerbeets1985b,Alba1986} The exponent
$n$ increases as $T$ approaches $T_{SG}$ from the low-$T$ side.  The
overall variation of $n$ below $T_{SG}$ is qualitatively the same as that
calculated for the Sherrington-Kirkpatrick model.\cite{Sherrington1975}

Figure \ref{fig08}(b) shows the $T$ dependence of the amplitude $A$ for
$t_{w} = 1.5 \times 10^{4}$ and $3.0 \times 10^{4}$ sec.  We find that $A$
is strongly dependent on $T$.  Irrespective of $t_{w}$, the amplitude $A$
shows two local maxima at 3 and 7.0 K. For comparison, we evaluate the $T$
dependence of $S_{max}^{0}$ [$= A(1-n)/e$], which corresponds to
Eq.(\ref{eq22}) with $m = 0$.  As the values of $n$ and $A$, here we use
the experimental values at each $T$ shown in Figs.~\ref{fig08}(a) and (b). 
We find that the $T$ dependence of $S_{max}^{0}$ thus calculated is in
excellent agreement with that of $S_{max}$ (see Fig.~\ref{fig07})
experimentally determined from the data of $S(t)$ vs $t$.

\subsection{\label{resultC}$S(t)$ under the $H$-shift}

\begin{figure}
\includegraphics[width=7.5cm]{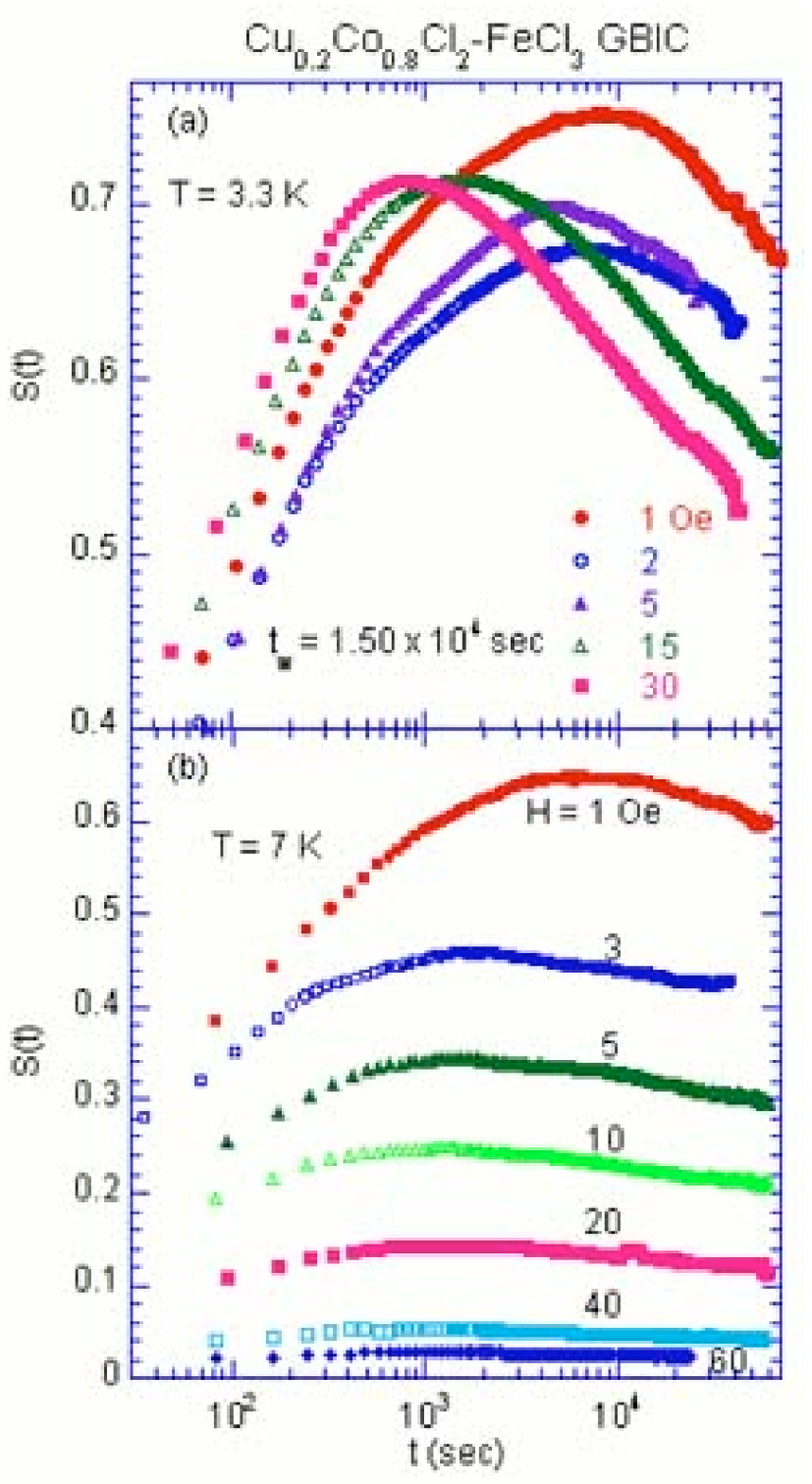}%
\caption{\label{fig09}(Color online)$S$ vs $t$ at various $H$.  (a) $T =
3.3$ K and $t_{w} = 1.5 \times 10^{4}$ sec.  (b) $T = 7.0$ K and $t_{w} =
1.5 \times 10^{4}$ sec.}
\end{figure}

\begin{figure}
\includegraphics[width=7.5cm]{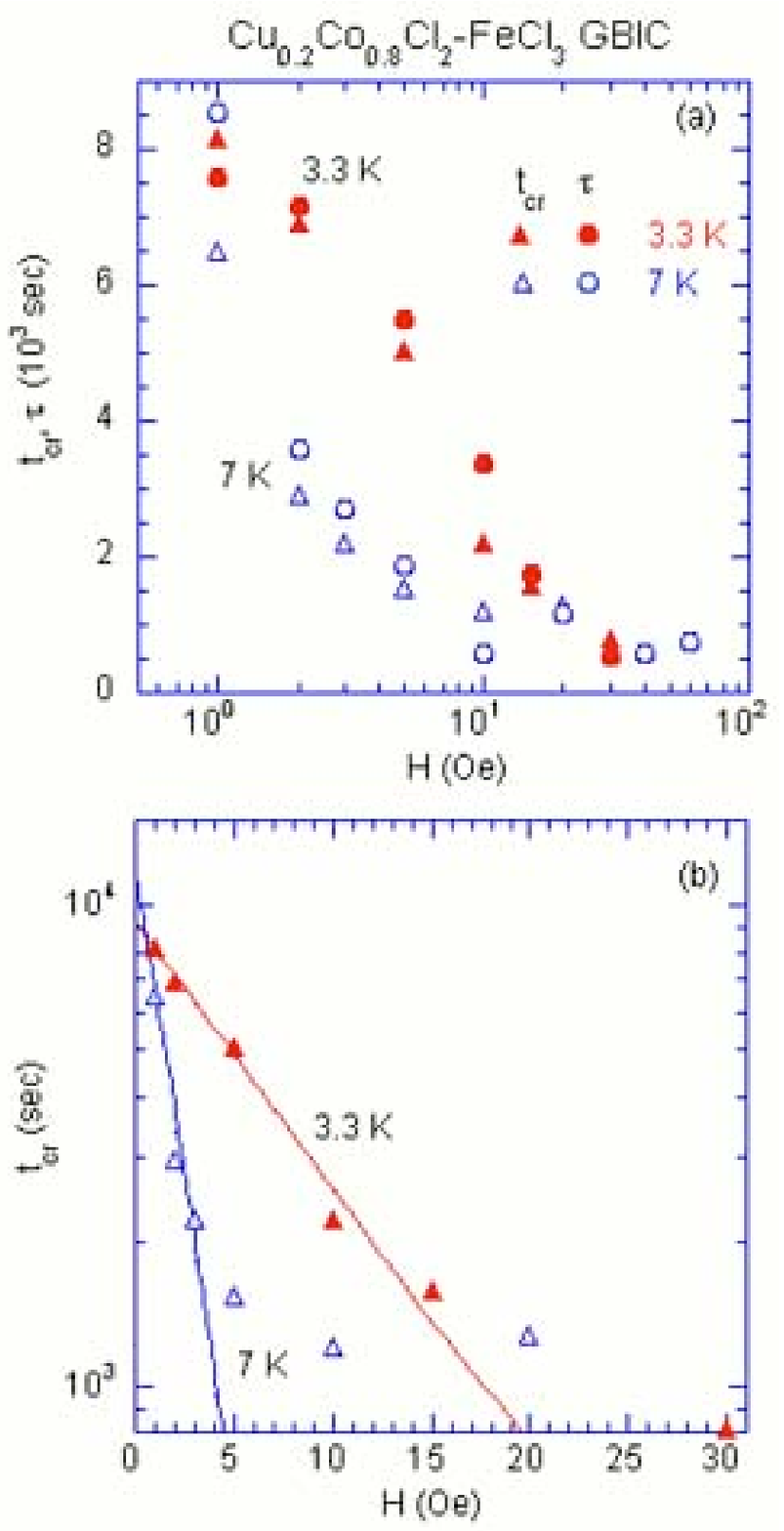}%
\caption{\label{fig10}(Color online)(a) $t_{cr}$ vs $H$ and $\tau$ vs $H$
at $T = 3.3$ and 7.0 K. The soild lines are guides to the eyes.  (a)
$t_{cr}$ vs $H$ at $T$ 3.3 and 7.0 K. The solid lines are least-squares
fits to Eq.(\ref{eq25}) for data at low $H$.  The fitting parameters are
given in the text.}
\end{figure}

Figures \ref{fig09}(a) and (b) show the $t$ dependence of $S(t)$ at various
$H$, where $T$ = 3.3 and 7.0 K, and $t_{w} = 1.5 \times 10^{4}$ sec.  The
relaxation rate $S(t)$ shows a peak at $t_{cr}$.  This peak greatly shifts
to the short-$t$ side with increasing $H$.  In Fig.~\ref{fig10}(a) we make
a plot of $t_{cr}$ and $\tau$ as a function of $H$ at $T$ = 3.3 and 7.0 K,
where $\tau$ is determined from the least squares fit of the data of
$\chi_{ZFC}$ vs $t$ to the stretched exponential relaxation with $m = 0$. 
The value of $t_{cr}$ is almost the same as that of $\tau$ at each $T$ and
$H$, indicating that the exponent $m$ is equal to zero (see
Sec.~\ref{backB}).  The time $t_{cr}$ undergoes a dramatic decrease at low
fields : $H \approx$ 2 - 3 Oe at $T = 7.0$ K and $H = 10$ Oe at $T = 3.3$
K. Note that the decrease of $t_{cr}$ with $H$ has been reported for the
ZFC relaxation in the RSG phase of the reentrant ferromagnet
Fe$_{0.3}$Al$_{0.7}$.\cite{Motoya2003} The increase of $n$ and $1/\tau$
with increasing $H$ has been reported for the TRM decay in the SG systems
such as Ag:[Mn (2.6 at.~\%) + Sb (0.46 at.~\%)]\cite{Hoogerbeets1985b} and
Cu: Mn (6 at.~\%).\cite{Chu1995}

The shift of $t_{cr}$ for $S(t)$ under the $H$-shift process is governed by
the mean domain-size $L_{T}(t)$ and overlap lengths $L_{H}$ defined
by\cite{Fisher1988}
\begin{equation}
L_{T}(t)/L_{0} \approx (t/t_{0})^{1/z(T)}
\text{  and  }
L_{H}/L_{0} \approx (H/\Upsilon_{H})^{-1/\delta},
    \label{eq24}
\end{equation}
respectively, where $z(T)$ and $\delta$ are the corresponding exponents for
$L_{T}(t)$ and $L_{H}$, and $\Upsilon_{H}$ is the magnetic field
corresponding to a wall stiffness $\Upsilon$ (a typical energy setting the
scale of free energy barriers between conformations).  Using the scaling
concept, the shift of $t_{cr}$ for $S(t)$ under the $H$-shift can be
approximated by\cite{Suzuki2003,Takayama2003}
\begin{equation}
\ln (t_{cr}/t_{w}) =-\alpha_{H}H,
    \label{eq25}
\end{equation}
in the limit of $H \approx 0$, where $\alpha_{H}$ is constant proportional
to $z(T)$ $t_{w}^{\delta /z(T)}$.  In Fig.~\ref{fig10}(b) we make a plot of
$t_{cr}$ vs $H$ at $T = 3.3$ and 7.0 K. We find that the data of $t_{cr}$
vs $H$ is well described by Eq.(\ref{eq25}) with $\alpha_{H} = 0.128 \pm
0.009$ and $t_{w} =(9.19 \pm 0.31) \times 10^{3}$ sec for $H<15$ Oe at $T =
3.3$ K and $\alpha_{H} = 0.63 \pm 0.15$ and $t_{w} =(12.0 \pm 2.6) \times
10^{3}$ sec for $H<1$ Oe at $T = 7.0$ K.

\begin{figure}
\includegraphics[width=7.5cm]{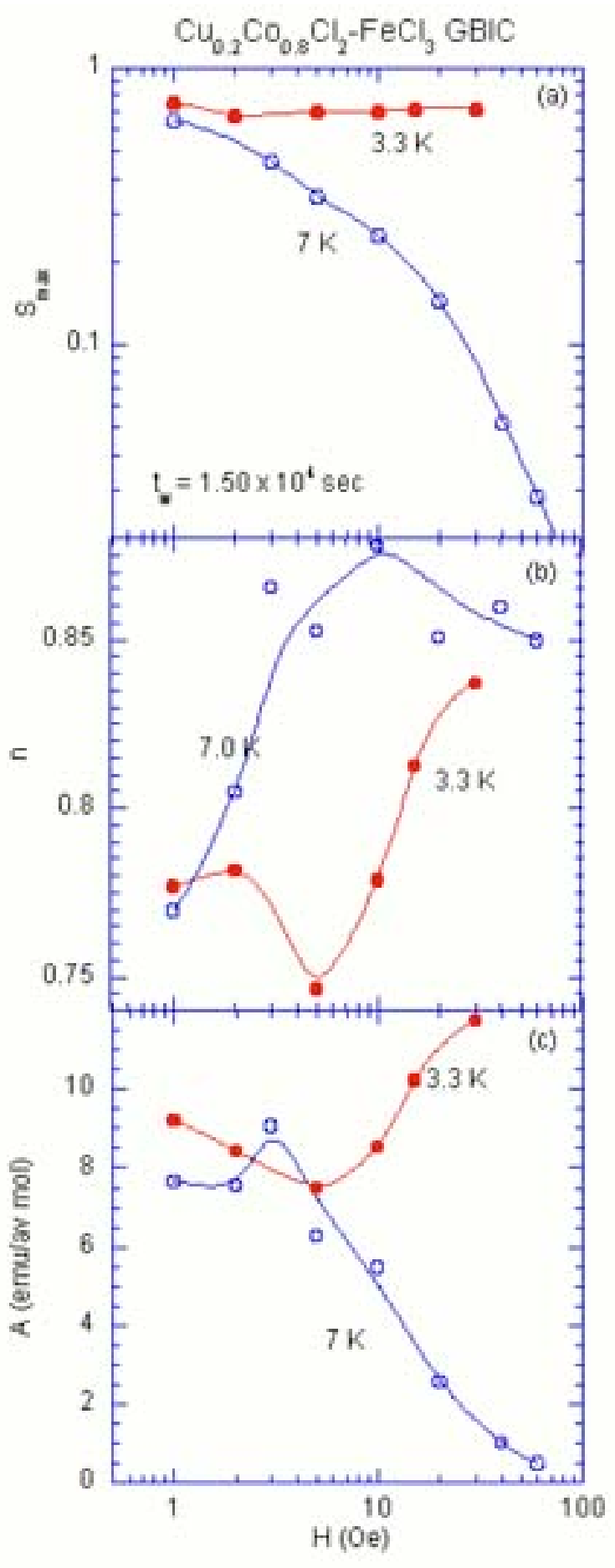}%
\caption{\label{fig11}(Color online)(a) $S_{max}$ vs $H$, (b) $n$ vs $H$,
and (c) $A$ vs $H$ at $T = 3.3$ and 7.0 K. $t_{w} = 1.5 \times 10^{4}$ sec. 
$H = 1$ Oe.  The soild lines are guides to the eyes.}
\end{figure}

In Fig.~\ref{fig11}(a) we show the $H$ dependence of $S_{max}$ at $T = 3.3$
and 7.0 K, where $t_{w} = 1.5 \times 10^{4}$ sec.  The value of $S_{max}$
is almost independent of $H$ for $0 \leq H \leq 40$ Oe.  In contrast, the
value of $S_{max}$ at 7.0 K greatly reduces with increasing $H$.  This
result agrees well with that reported by Jonason and Nordblad\cite{Jonason1998}
with the gradual decrease of $S_{max}$ at 16 K with $H$ and the dramatic
decrease in $S_{max}$ at 30 K with $H$ at weak fields less than 0.2 Oe. 
Figure \ref{fig11}(b) shows the $H$ dependence of the exponent $n$ at $T =
3.3$ and 7.0 K, where $t_{w} = 1.5 \times 10^{4}$ sec.  The $H$ dependence
of $n$ at 7.0 K is rather different from that at 3.3 K at low $H$.  The
value of $n$ at $T = 3.3$ K has a local minimum at $H = 5$ Oe, while the
value of $n$ at $T = 7.0$ K has a local maximum around 10 Oe.  The value of
$n$ tends to saturate to 0.82 - 0.84 at $H$ higher than 30 Oe, independent
of $T$.  Figure \ref{fig11}(c) shows the $H$ dependence of $A$ at $T = 3.3$
and 7.0 K, where $t_{w} = 1.5 \times 10^{4}$ sec.  The amplitude $A$ at 7.0
K deviates from that at 3.3 K above 5 Oe, indicating the fragility of the
aging state in the FM phase against a weak magnetic field-perturbation. 
For comparison, we make a plot of $S_{max}^{0}$ [$=A(1-n)/e$] as a function of $H$, where
$A$ and $n$ experimentally determined are used.  There is a very good
agreement between $S_{max}$ (see Fig.~\ref{fig11}(a)) and $S_{max}^{0}$.

\subsection{\label{resultD}Effect of $t_{w}$ on $S(t)$}

\begin{figure}
\includegraphics[width=7.5cm]{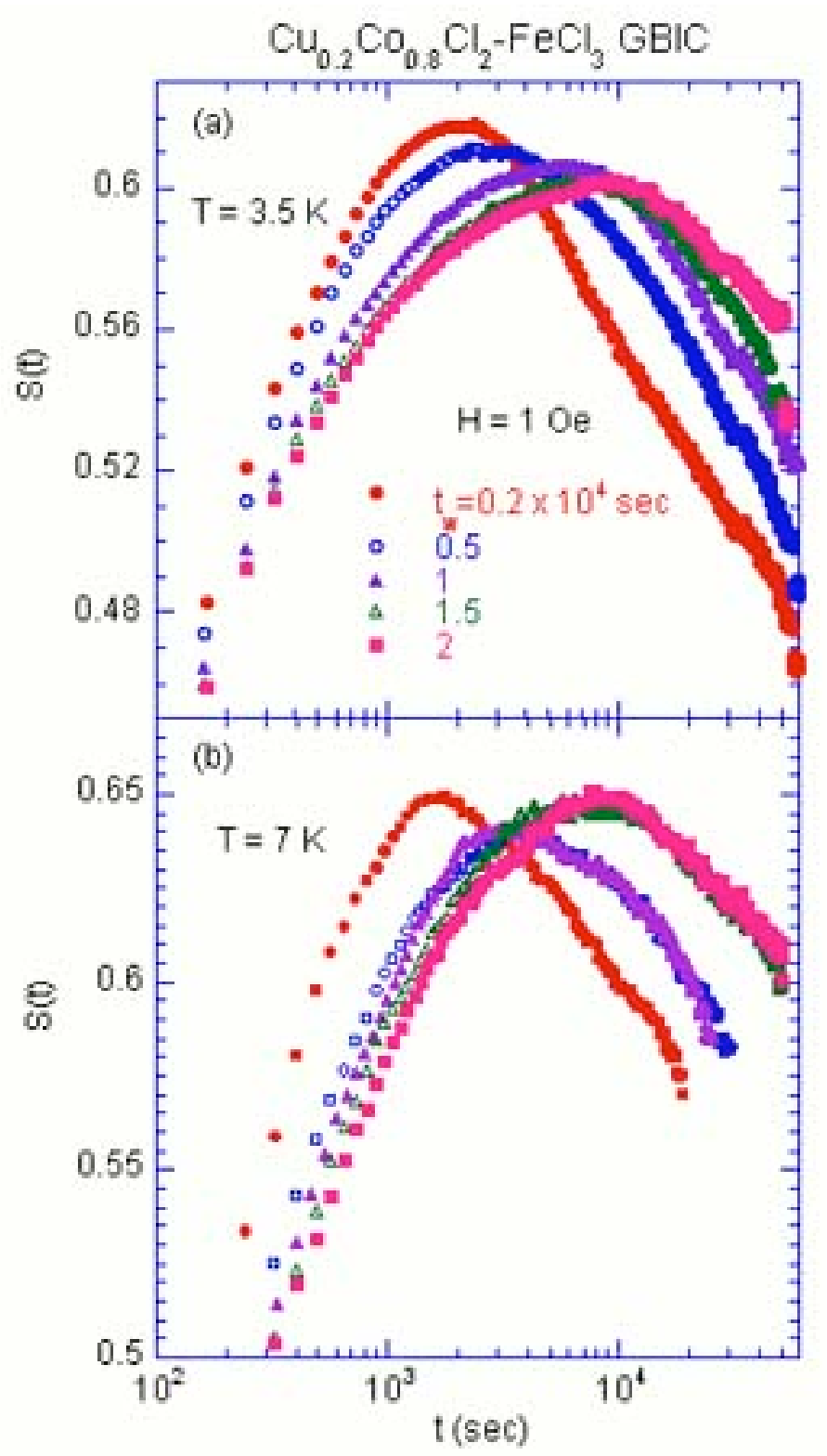}%
\caption{\label{fig12}(Color online)$S$ vs $t$ at various $t_{w}$ ($2
\times 10^{3} \leq t_{w} \leq 3.0 \times 10^{4}$ sec).  $H = 1$ Oe.  (a) $T
= 3.5$ K. (b) $T = 7.0$ K.}
\end{figure}

\begin{figure}
\includegraphics[width=7.5cm]{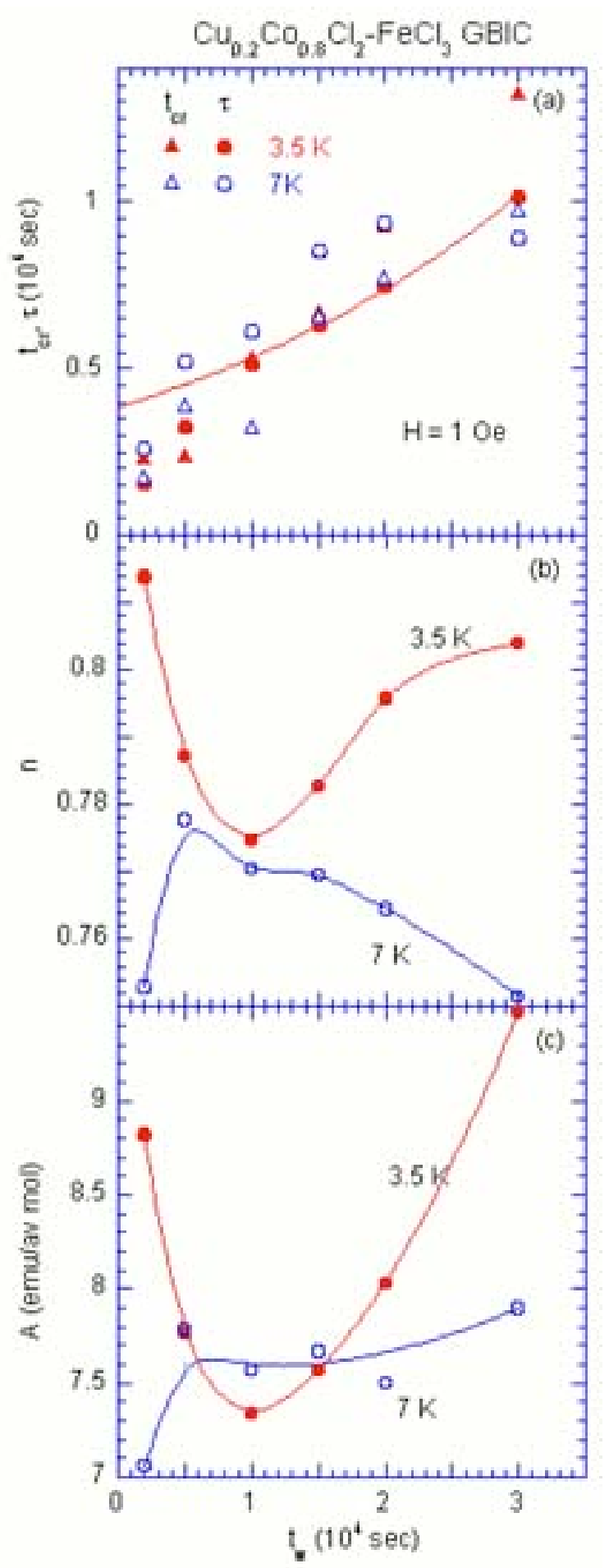}%
\caption{\label{fig13}(Color online)(a) $t_{cr}$ vs $t_{w}$ and $\tau$ vs
$t_{w}$.  The solid line is a least-squares fit to $\tau =
\tau_{0}\exp(t_{w}/t_{0})$ for $\tau$ at 3.5 K ($t_{w} \geq 1.0 \times
10^{4}$ sec).  The fitting parameters are given in the text.  (b) $n$ vs
$t_{w}$, and (c) $A$ vs $t_{w}$ at $T = 3.5$ and 7.0 K. The soild lines are
guides to the eyes.}
\end{figure}

Figure \ref{fig12} shows the $t$ dependence of $S(t)$ at various $t_{w}$
for $T = 3.5$ and 7.0 K, where $H = 1$ Oe.  The relaxation rate $S(t)$
shows a peak at $t_{cr}$.  This peak shifts to long-$t$ side with
increasing $t_{w}$.  The least squares fit of the data of $\chi_{ZFC}(t)$
vs $t$ for $1.0 \times 10^{2} \leq t \leq 6.0 \times 10^{4}$ sec to Eq.(\ref{eq16})
yields the exponent $n$, the relaxation time $\tau$, and amplitude $A$ for
the stretched exponential relaxation.  Figure \ref{fig13}(a) shows the
$t_{w}$ dependence of $t_{cr}$ and $\tau$ at $T = 3.5$ and 7.0 K. At $T =
7.0$ K, $t_{cr}$ is nearly equal to $\tau$ for $0<t_{w} \leq 3.0 \times
10^{4}$ sec, while at $T = 3.5$ K $t_{cr}$ is much longer than $\tau$ for
$t_{w} \geq 1.5 \times 10^{4}$ sec.  Both $t_{cr}$ and $\tau$ increases with
increasing $t_{w.}$ It has been reported that $\tau$ varies exponentially
with $t_{w}$ as $\tau = \tau_{0}\exp(t_{w}/t_{0})$ for Ag: Mn (2.6
at.~\%).\cite{Hoogerbeets1985b} Our data of $\tau$ vs $t_{w}$ at 3.5 K are
well fitted to this equation in spite of the limited $t_{w}$ region ($1.0
\times 10^{4} \leq t_{w} \leq 3.0 \times 10^{4}$ sec); $\tau_{0} = (3.83
\pm 0.012) \times 10^{3}$ sec and $t_{0} = (3.07 \pm 0.13) \times 10^{4}$
sec.  Figure \ref{fig13}(b) shows the $t_{w}$ dependence of $n$ at $T =
3.5$ and 7.0 K. The exponent $n$ at $T = 3.5$ K shows a local minimum at
$t_{w} = 1.0 \times 10^{4}$ sec and increases with further increasing
$t_{w}$.  In contrast, the exponent $n$ at 7.0 K is smaller than that at
3.5 K for any $t_{w}$.  It exhibits a local maximum around $t_{w} = 5.0
\times 10^{3}$ sec.  Similar local minimum in $n$ vs $t_{w}$ is observed in
the SG phase of the spin glass Cr$_{83}$Fe$_{17}$ [$T_{SG} = 13$
K]:\cite{Mitchler1993} $n$ at $T = 8$ K shows a local minimum around $t_{w}
= 3.0 \times 10^{3}$ sec and increases with further increasing $t_{w}$.  In
contrast, there is no local minimum in $n$ vs $t_{w}$ in the RSG phase of
(Fe$_{0.65}$Ni$_{0.35}$)$_{0.882}$Mn$_{0.118}$\cite{Li1994} ($n$ increases
with increasing $t_{w}$) and in the SG phase of Ag: Mn (2.6 and 4.1
at.~\%)\cite{Hoogerbeets1985b} ($n$ is independent of $t_{w}$).  Figure
\ref{fig13}(c) shows the $t_{w}$ dependence of the amplitude $A$ at $T =
3.5$ and 7.0 K. The amplitude $A$ at 3.5 K shows a local minimum around
$t_{w} = 1.0 \times 10^{4}$ sec, while $A$ at 7.0 K is almost independent
of $t_{w}$.  In summary, so far there has been no theory to explain the
$t_{w}$ dependence of $n$, $\tau$, $t_{cr}$ and $A$ in both the RSG and FM
phases.  However, our results suggest that the RSG and FM phases are
essentially nonequilibrium phases.  The nature of the aging behavior in the
FM phase is rather different from that in the RSG phase.

\subsection{\label{resultE}$S(t)$ under the $T$-shift}

\begin{figure}
\includegraphics[width=7.5cm]{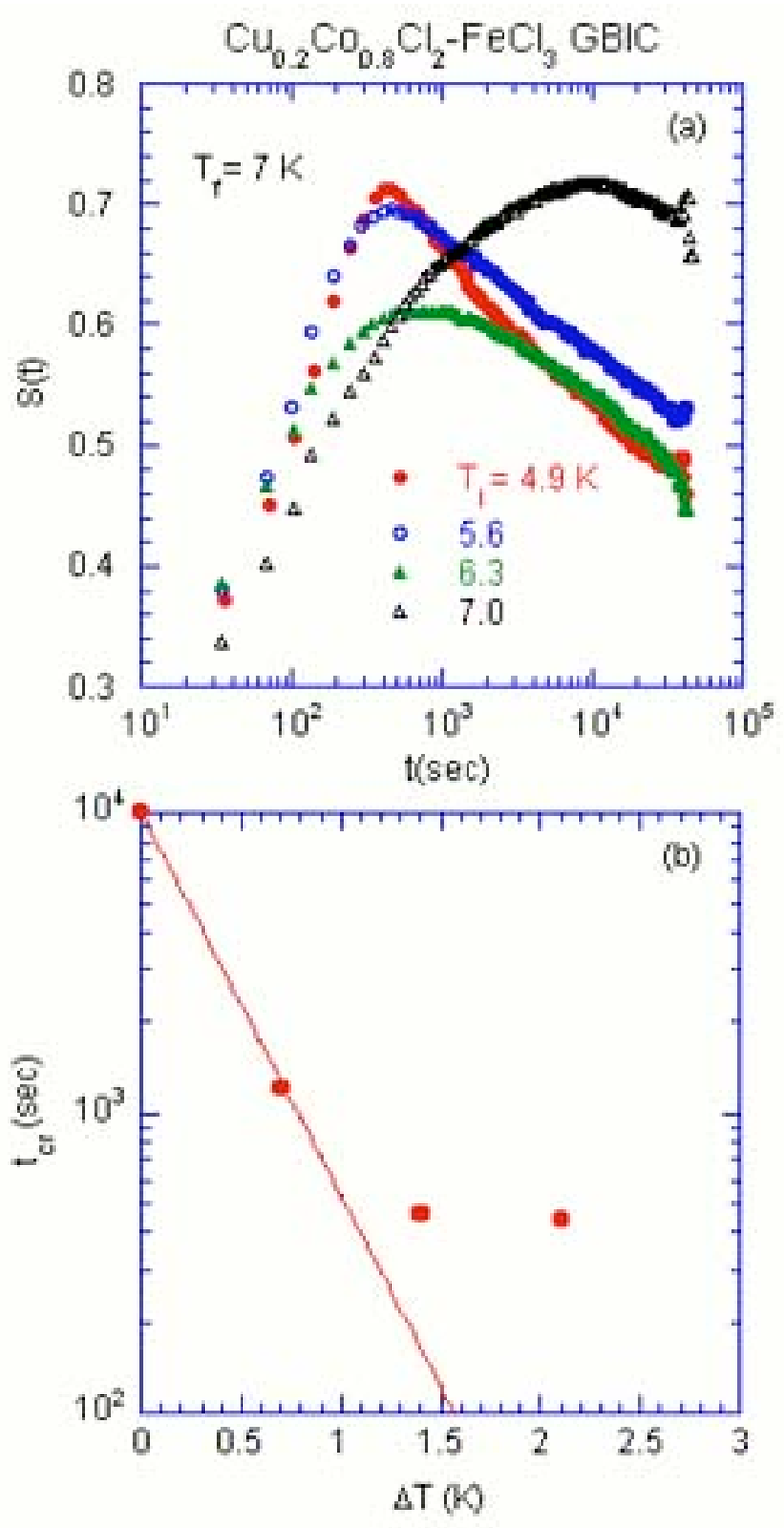}%
\caption{\label{fig14}(Color online)(a) $t$ dependence of $S$ at $T = T_{f}
= 7.0$ K under the $T$ shift from $T_{i}$ to $T_{f}$, where $T_{i}$ = 4.9,
5.6, and 6.3 K. $H = 1$ Oe.  The ZFC aging protocol is as follows:
quenching of the system from 50 K to $T_{i}$, and isothermal aging at $T =
T_{i}$ and $H = 0$ for $t_{w} = 3.0 \times 10^{4}$ sec.  The measurement
was started at $t = 0$, just after $T$ was shifted from $T_{i}$ to $T_{f}$
and subsequently $H$ (= 1 Oe) was turned on.  (b) $t_{cr}$ vs $\Delta T$. 
The solid line is a fitting curve to Eq.(\ref{eq27}).  The fitting
parameters are given in the text.}
\end{figure}

Figure \ref{fig14}(a) show the relaxation rate $S(t)$ measured at $T_{f} =
7.0$ K for different initial temperature $T_{i}$ (= 4.9, 5.6, and 6.3 K),
where $t_{w} = 3.0 \times 10^{4}$ sec and $H = 1$ Oe.  The measurement was
carried out as follows.  After the ZFC protocol from 50 K to $T_{i}$ below
$T_{c}$, the system was kept at $T_{i}$ for a wait time $t_{w}$. 
Immediately after $T$ was changed from $T_{i}$ to $T_{f}$ (the positive $T$
shift; $\Delta T>0$ where $\Delta T = T_{f} - T_{i}$), the magnetic field
was applied and the $t$ dependence of $\chi_{ZFC}$ was measured.

With increasing $\Delta T$, the time $t_{cr}$ shifts to the low-$t$ side. 
The relaxation rate $S(t)$ has a peak at $t_{cr} = 440$ - 460 sec at
$\Delta T = 2.1$ K and 1.4 K, suggesting a rejuvenation of the system
during the positive $T$-shift.  For $\Delta T = 0.7$ K two peaks appear at
570 sec and 1210 sec, respectively, suggesting a coexistence of two
characteristic ages in the system due to the partial rejuvenation and
original aging at $T_{i}$ for $t_{w} = 3.0 \times 10^{4}$ sec.  This
behavior is explained in terms of the overlap length $L_{\Delta T}$, which
becomes smaller as $\Delta T$ becomes large.  When $R_{Ti}(t_{w}) \ll
L_{\Delta T}$ corresponding to the case of $\Delta T<\Delta T_{thresold}$,
no initialization occurs in the system.  There is only one type of domain. 
Thus $S(t)$ exhibits a peak around $t_{w}$.  For $R_{Ti}(t_{w}) \gg
L_{\Delta T}$ corresponding to the case of $\Delta T>\Delta T_{thresold}$,
some domains fractures into smaller domains of dimension $L_{\Delta T}$. 
Thus $S(t)$ has two peaks at $t_{cr}$ ($\approx t_{w}$) and a time
much shorter than $t_{cr}$.

The shift of $t_{cr}$ for $S(t)$ under the $T$-shift aging process is
governed by the mean domain-size $L_{T}(t)$ and overlap length $L_{\Delta
T}$ defined by\cite{Fisher1988}
\begin{equation}
L_{\Delta T}/L_{0} \approx 
(T^{1/2}\mid \Delta T \mid /\Upsilon_{T}^{3/2})^{-1/\zeta},
    \label{eq26}
\end{equation}
where $\zeta$ is the corresponding exponent and $\Upsilon_{T}$ is the
temperature corresponding to the wall stiffness $\Upsilon$.  Using the
scaling concepts, the shift of $t_{cr}$ for $S(t)$ under the $T$-shift can
be approximated by\cite{Suzuki2003,Jonsson2002,Jonsso2003}
\begin{equation}
\ln (t_{cr}/t_{w})=-\alpha_{T} \mid \Delta T \mid,
    \label{eq27}
\end{equation}
in the limit of $H \approx 0$, where $\alpha_{T}$ is a constant
proportional to $z(T_{f})T_{f}^{1/2}t_{w}^{\zeta/z(T_{i})}$.  In
Fig.~\ref{fig14}(b) we make a plot of $t_{cr}$ vs $\Delta T$ at $T_{f} =
7.0$ K. We find that the data of $t_{cr}$ vs $\Delta T$ is described by
Eq.(\ref{eq27}) with $\alpha_{T} = 2.93 \pm 0.33$ and $t_{w} = (10.0 \pm
0.3) \times 10^{3}$ sec for $\Delta T<0.7$ K.

Similar behaviors in $S(t)$ under the $T$ shift have been observed in other
reentrant ferromagnets.  In the FM phase of
(Fe$_{0.20}$Ni$_{0.80}$)$_{75}$P$_{16}$B$_{6}$Al$_{3}$,\cite{Jonason1996a,Jonason1999}
the amplitude of the maximum in $S(t)$ at $t$ $\approx t_{w}$ decreases
with increasing $\mid \Delta T \mid$ (both the positive and negative
$T$-shift) and a second maximum gradually develops at a shorter time,
indicating a partial rejuvenation.  In the RSG phase of
(Fe$_{0.65}$Ni$_{0.35}$)$_{0.882}$Mn$_{0.118}$ double peaks in $S(t)$ vs
$t$ are observed under a temperature-cycling experiment:\cite{Li1994} an
initial wait time $t_{w}$ ($= 1.0 \times 10^{4}$ sec) at $T$, followed by a
temperature cycle $T \rightarrow T+\Delta T \rightarrow T$ of the duration
$t_{cycle}$ (= 300 sec).

\subsection{\label{resultF}$\chi^{\prime\prime}(\omega,t)$}

\begin{figure}
\includegraphics[width=7.5cm]{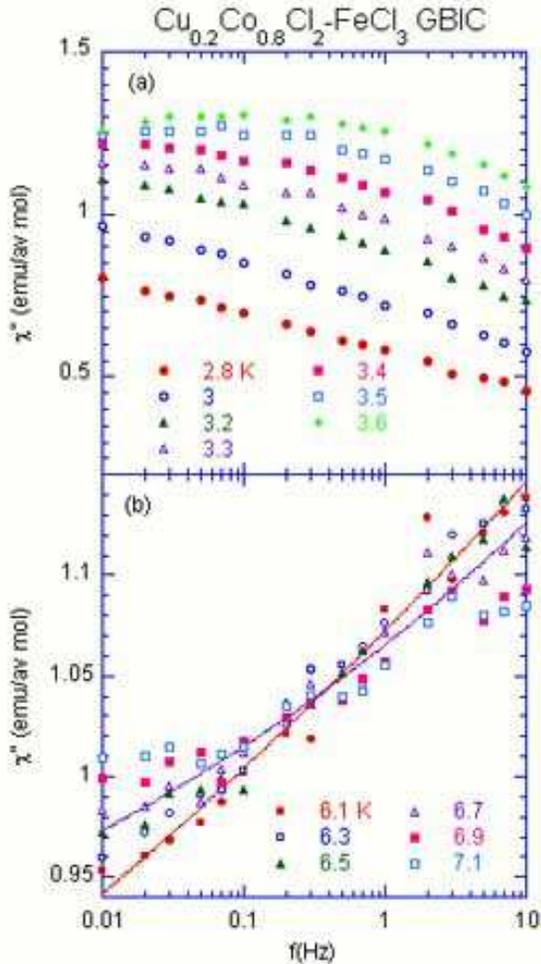}%
\caption{\label{fig15}(Color online)(a) and (b) $f$ dependence of
$\chi^{\prime\prime}(\omega,T)$ at various $T$.  $h = 50$ mOe.  The solid
lines are least-squares fits to Eq.(\ref{eq15}) for $T = 6.1$ and 6.7 K.
The fitting parameters are given in the text.}
\end{figure}

In previous paper\cite{Suzuki2004} we have reported the $f$ dependence of
the equilibrium absorption $\chi^{\prime\prime}(\omega,T)$ at various $T$
in the vicinity of $T_{RSG}$, respectively.  The absorption
$\chi^{\prime\prime}(\omega,T)$ curves exhibit different characteristics
depending on $T$.  Above $T_{RSG}$, $\chi^{\prime\prime}(\omega,T)$ shows a
peak at a characteristic frequency, shifting to the low $f$-side as $T$
decreases.  Below $T_{RSG}$, $\chi^{\prime\prime}(\omega,T)$ shows no peak
for $f \geq 0.01$ Hz.  In Figs.~\ref{fig15} (a) and (b), for comparison we
show the $f$ dependence of $\chi^{\prime\prime}(\omega,T)$ at fixed $T$
($2.8 \leq T \leq 3.6$ K and $6.1 \leq T \leq 7.1$ K).  The absorption
$\chi^{\prime\prime}(\omega,T)$ decreases with increasing $f$ in the RSG
phase, while it increases with increasing $f$ in the FM phase.  The $f$
dependence of $\chi_{eq}^{\prime\prime}(\omega,T)$ can be expressed by a
power law form ($\approx \omega^{\alpha}$), where the magnitude of the
exponent $\alpha$ is very small.  The sign of $\alpha$ may be negative for
the RSG phase and is positive for the FM phase.  According to the
fluctuation-dissipation theorem, the magnetic fluctuation spectrum
$P(\omega,T)$ is related to $\chi_{eq}^{\prime\prime}(\omega,T)$ by
$P(\omega,T) = 2k_{B}T\chi_{eq}^{\prime\prime}(\omega,T)/\omega$.  Then
$P(\omega,T)$ is proportional to $\omega^{-1+\alpha}$, which is similar to
$1/\omega$ character of typical SG systems.  Note that for the spin
glass CdCr$_{1.7}$In$_{0.3}$S$_{4}$ ($T_{SG} = 16.7$ K),\cite{Hammann1990}
$\alpha$ ($\approx 0.1$) is positive at $T_{SG}$ and decreases with
decreasing $T$.  The sign of $\alpha$ changes from positive to negative
below 5 K: $\alpha \approx -0.03$.

In contrast, $\chi^{\prime\prime}(\omega,T)$ for $6.0 \leq T \leq 7.4$ K
increases with increasing $f$.  The least squares fit of the data of
$\chi^{\prime\prime}(\omega,T)$ vs $f$ for $0.01 \leq f\leq 10$ Hz to the
power law form ($\approx \omega^{\alpha}$) yields the value of $\alpha$. 
The exponent $\alpha$ increase with increasing $T$ for $6.1 \leq T \leq
7.2$ K: $\alpha = 0.04 \pm 0.01$ at 6.1 K, $0.081 \pm 0.02$ at 6.7 K and
$\alpha = 0.12 \pm 0.02$ at 7.2 K.

\begin{figure}
\includegraphics[width=7.5cm]{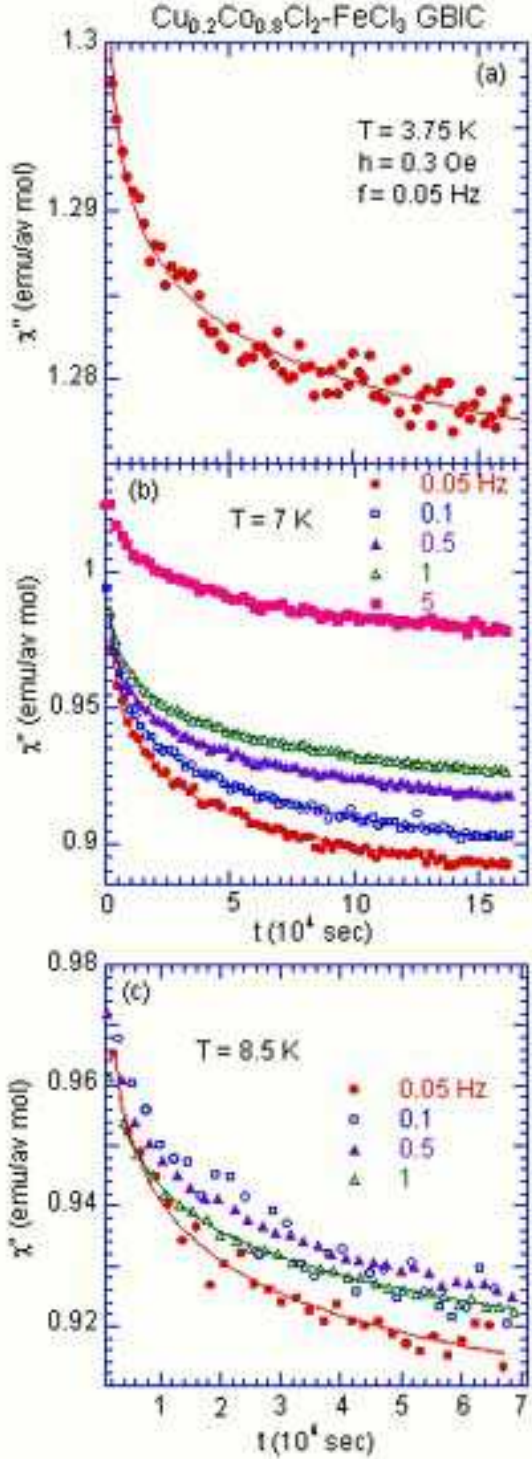}%
\caption{\label{fig16}(Color online)$t$ dependence of
$\chi^{\prime\prime}(\omega,t)$ at (a) 3.75 K, (b) $T = 7.0$ K, and (c) $T
= 8.5$ K. $f$ = 0.05, 0.1, 0.5 and 1 Hz.  The time $t$ is the time taken
after the sample is quenched from 50 K to $T$.  $H = 0$.  The solid lines
are the least-squares fits to Eq.(\ref{eq28}).  The fitting parameters are
listed in Table \ref{table01}.}
\end{figure}

\begin{table}
\caption{\label{table01}Exponents $\alpha$ (or $b$) determined from the least squares
fits of $\chi^{\prime\prime}(\omega,t)$ at $f$ to the power law form by
given by Eq.(\ref{eq28}).  $T$ = 3.5,
3.75, 7 and 8.5 K. The parameters with the suffix ``c'' denote those
determined from the least squares fits of the data to a power-law form
$(t+t_{0})^{-b}$ instead of $t^{-b}$, where $t_{0}$ is regarded as a
fitting parameter.}
\begin{ruledtabular}
\begin{tabular}{lllll}
$T$(K) & $f$(Hz) & $\alpha (b)$ & $A$ & $\chi_{0}^{\prime\prime}(\omega)$\\
\hline
3.5 & 0.1 & 0.070 & 0.182 & 1.100\\
3.75 & 0.05 & 0.084 & 0.137 & 1.228\\
7 & 0.05 & 0.074 & 0.334 & 0.662\\
7 & 0.1 & 0.045 & 0.458 & 0.537\\
7 & 0.5 & 0.042 & 0.386 & 0.606\\
7 & 1 & 0.029 & 0.505 & 0.489\\
7 & 5 & 0.015 & 0.770 & 0.265\\
7$^{c}$ & 0.05 & 0.224 & 0.216 & 0.822\\
7$^{c}$ & 0.1 & 0.072 & 0.324 & 0.676\\
7$^{c}$ & 0.5 & 0.069 & 0.267 & 0.729\\
7$^{c}$ & 1 & 0.047 & 0.333 & 0.665\\
8.5 & 0.05 & 0.147 & 0.148 & 0.835\\
8.5 & 1 & 0.046 & 0.268 & 0.702\\
\end{tabular}
\end{ruledtabular}
\end{table}

We have also measured the $t$ dependence of $\chi^{\prime\prime}(\omega,t)$
at $T$ = 3.3, 3.5, 3.75, 7 and 8.5 K, where $H = 0$, $h = 0.3$ Oe, and $f$
= 0.05 - 5 Hz.  The system was quenched from 100 K to $T$ at time (age)
zero.  The origin $t = 0$ is a time at which the temperature becomes stable
at $T$ within $\pm 0.01$ K. The change of $\chi^{\prime\prime}(\omega,t)$
with $t$ below $T_{RSG}$ is not so prominent compared to that above
$T_{RSG}$, partly because of relatively small magnitude of
$\chi^{\prime\prime}$ below $T_{RSG}$.  In Figs.~\ref{fig16}(a)-(c), we
shows the $t$ dependence of $\chi^{\prime\prime}(\omega,t)$ at various $f$,
where $T$ = 3.75, 7.0, and 8.5 K. We find that the decrease of
$\chi^{\prime\prime}$ with increasing $t$ is well described by a power-law
decay of the quasi-equilibrium part
\begin{equation}
\chi^{\prime\prime}(\omega,t) \approx 
\chi_{eq}^{\prime\prime}(t) =\chi^{\prime\prime}_{0}(\omega)
+A(\omega)t^{-\alpha}
    \label{eq28}
\end{equation}
where $\chi^{\prime\prime}_{0}(\omega)$ and $A(\omega)$ are $t$-independent
constants.  It is predicted that $\chi_{eq}^{\prime\prime}(t)$ exhibits no
$\omega t$-scaling law [$\chi_{eq}^{\prime\prime}(t)$ $\approx
t^{-\alpha}$].  In the limit of $t \rightarrow \infty$,
$\chi^{\prime\prime}(\omega,t)$ tends to $\chi^{\prime\prime}_{0}(\omega)$
which is assumed to be equal to $\chi^{\prime\prime}(\omega,T)$.  The least
squares fit of the data of $\chi^{\prime\prime}(\omega,t)$ at each $T$ to
Eq.(\ref{eq28}) yields parameters listed in Table \ref{table01} at $T$ = 3.5,
3.75, 7, and 8.5 K. The exponent $\alpha$ below $T_{RSG}$ is nearly equal to
0.07 - 0.084.  At 7.0 K, the exponent $\alpha$ is dependent on $f$ :
$\alpha = 0.074$ at $f = 0.05$ Hz and $\alpha = 0.015$ at $f = 5$ Hz.  The
value of $\chi^{\prime\prime}_{0}$ at $T = 7.0$ K tends to decrease with
increasing $f$.  This seems to be inconsistent with our result that
$\chi^{\prime\prime}_{eq}$ increases with increasing $f$.  Since the value
of $A$ tends to increase with increasing $f$, it follows that the second
term of Eq.(\ref{eq28}) does not obey the $\omega t$-scaling law as is
predicted.

In general, $\chi^{\prime\prime}(\omega,t)$ consists of the
quasi-equilibrium part $\chi_{eq}^{\prime\prime}(t)$ and the aging
relaxation part $\chi_{ag}^{\prime\prime}(\omega t)$.  The aging part
$\chi_{ag}^{\prime\prime}(\omega t)$ is predicted to be described by a
$\omega t$-scaling law [$\chi_{ag}^{\prime\prime}(\omega t) \approx (\omega
t)^{-b}$].\cite{Komori1999} Note that the $\omega t$-scaling of
$\chi^{\prime\prime}(\omega,t)$ is observed in the RSG phase ($T = 8$ K)
and the FM phase ($T = 40$ K) of the reentrant ferromagnet
CdCr$_{1.9}$In$_{0.1}$S$_{4}$ ($T_{RSG} = 18$ K and $T_{c} = 50$
K).\cite{Vincent2000a} The absorption $\chi^{\prime\prime}(\omega,t)$
mainly arises from $\chi_{ag}^{\prime\prime}(\omega t)$ but not from
$\chi_{eq}^{\prime\prime}(t)$.  The aging relaxation part
$\chi_{ag}^{\prime\prime}(\omega,t)$ is assumed to be described by the
power-law form [$\omega (t+t_{0})]^{-b}$, instead of $(\omega t)^{-b}$,
where $b = 0.2$ for both temperatures and $t_{0}$ is an off-set time which
takes into account the fact that the cooling procedure is not
instantaneous.  The least squares fit of our data of
$\chi^{\prime\prime}(\omega,t)$ vs $t$ at $T = 7.0$ K to the power-law form
$A(t+t_{0})^{-b}$ yields the parameters listed in Table \ref{table01}, where
$t_{0}$ is one of the fitting parameters.  We find that the value of $b$
for $f = 0.05$ Hz ($b = 0.22 \pm 0.06$) is close to that for
CdCr$_{1.9}$In$_{0.1}$S$_{4}$.\cite{Vincent2000a} However, the value of $b$
for higher $f$ is still on the order of 0.047 - 0.07, which is much smaller
than 0.2.  This result indicates that $\chi^{\prime\prime}(\omega, t)
\approx \chi_{eq}^{\prime}(t) \approx t^{-\alpha}$ in our system.

\section{\label{dis}DISCUSSION}
\subsection{\label{disA}Dynamic nature of the RSG and FM phases}
The nature of the RSG and FM phases for
Cu$_{0.2}$Co$_{0.8}$Cl$_{2}$-FeCl$_{3}$ GBIC is summarized as follows.  The
static and dynamic behavior of the RSG phase is characterized by that of
the normal SG phase: a critical exponent $\beta = 0.57 \pm 0.10$ for the SG
parameter and a dynamic critical exponent $x = 8.5 \pm 1.8$ for the
characteristic relaxation time.\cite{Suzuki2004} The aging phenomena are
clearly seen in the RSG phase, although no appreciable nonlinear magnetic
susceptibility is observed.  The exponent $n$ for the stretched exponential
relaxation exhibits a local minimum just below $T_{RSG}$, and increases
when $T$ approaches $T_{RSG}$ from the low-$T$ side.  The relaxation time
$\tau$ for the stretched exponential relaxation drastically increases with
decreasing $T$ below $T_{RSG}$.  In this sense, it follows that the RSG
phase below $T_{RSG}$ is a normal SG phase.

In contrast, the dynamic nature of the FM phase is rather different from
that of an ordinary ferromagnet.  A prominent nonlinear susceptibility is
observed between $T_{RSG}$ and $T_{c}$.\cite{Suzuki2004} Aging phenomena
and a partial rejuvenation effect under the $T$-shift are also seen in the
FM phase.  This aging state disappears even in a weak magnetic field
($H>H_{0}$; $H_{0} = 2$ - 5 Oe).  The relaxation time $\tau$ (or $t_{cr}
\approx \tau$) as a function of $T$ exhibits a local maximum between
$T_{RSG}$ and $T_{c}$.  These results suggest that no long range
ferromagnetic correlation exist in the RSG phase.  The chaotic behavior of
the FM phase is rather similar to that observed in the RSG phase.

Similar behaviors have been observed in the reentrant ferromagnets.  In
(Fe$_{0.20}$Ni$_{0.80}$)$_{75}$P$_{16}$B$_{6}$Al$_{3}$
\cite{Jonason1996a,Jonason1996b,Jonason1998,Jonason1999} the relaxation
time diverges at a finite temperature ($\approx T_{RSG}$) with a dynamic critical exponent
similar to that observed for normal SG transitions.  The FM phase
just above $T_{RSG}$ shows a dynamic behavior characterized by an aging
effect and chaotic nature similar to that of RSG phase.  In
CdCr$_{2x}$In$_{2(1-x)}$S$_{4}$ with $x$ = 0.90, 0.95, and
1.0,\cite{Vincent2000a,Vincent2000b} the aging behavior of the low
frequency AC susceptibility is observed both in the FM and RSG phases,
with the same qualitative features as in normal SG systems.

In the mean-field
picture,\cite{Sherrington1975,Edwards1975,Parisi1979,Toulouse1980} a true
reentrance from the FM phase to the normal SG phase is not predicted. 
There exists a normal FM long range order in the FM phase.  This picture,
which assumes infinite-range interactions, is not always appropriate for
real reentrant magnets where the short-range interactions are large and
random in sign and the spin symmetry is rather Heisenberg-like than
Ising-like.  In a picture proposed by Aeppli et al.\cite{Aeppli1983}, the
system in the FM phase consists of regions which by themselves would order
ferromgnetically and other regions forming paramagnetic (PM) clusters.  The
frustrated spins in the PM clusters can generate random molecular fields
which act on the unfrustrated spins in the infinite FM network.  In the FM
phase well above $T_{RSG}$, the fluctuations of the spins in the PM
clusters are so rapid that the FM network is less influenced by them and
their effect is only to reduce the net FM moment.  On decreasing the
temperature toward $T_{RSG}$, the thermal fluctuations of the spins in the
PM clusters become slower.  The coupling between the PM clusters and the FM
network becomes important and the molecular field from the slow PM spins
acts as a random magnetic field.  This causes a breakup of the FM network
into finite-sized clusters.  Below $T_{RSG}$, the
ferromagnetism completely disappears, leading to a RSG phase.

\subsection{\label{disB}Comparison with $S(t)$ in other reentrant ferromagnets}
We have shown that the $t$ dependence of $S(t)$ is strongly dependent on
$T$, $H$, and $t_{w}$.  Here we compare our results on $S(t)$ with those
observed in other reentrant ferromagnets.  In order to facilitate the
comparison, we have determined the time $t_{cr}$ and the peak height
$S_{max}$ from their original data of $S$ vs $t$.  The first case is the
results of $S(t)$ for
(Fe$_{0.20}$Ni$_{0.80}$)$_{75}$P$_{16}$B$_{6}$Al$_{3}$ ($T_{RSG} = 14.7$ K
and $T_{c} = 92$ K) reported by Jonason and Nordblad.\cite{Jonason1998}
They have shown that $S(t)$ exhibits a peak at $t_{cr}$ in both RSG 
and FM phases, indicating that the aging phenomena occur in both phases. 
These results are in good agreement with our results [see Figs.~\ref{fig03}
and \ref{fig04}].  The aging state in the FM phase is fragile against a
very weak magnetic-field perturbation.  There is a threshold magnetic field
$H_{0}$, below which there is a linear response of the relaxation.  The
field $H_{0}$ undergoes a dramatic decrease with increasing $T$ above
$T_{RSG}$.  The value of $H_{0}$ ($\approx 0.5$ Oe) for $T<T_{RSG}$ is much
lower than that for our system ($H_{0} \approx$ 2 - 5 Oe).  The time
$t_{cr}$ decreases with increasing $T$ at low $T$, showing a local minimum
around 23 K, and increases with increasing $T$ between 25 and 30 K.
Although there have been no data on $S(t)$ above 30 K, it is assumed that
$t_{cr}$ shows a local maximum between 30 K and $T_{c}$, since $t_{cr}$
should reduce to zero above $T_{c}$.  Such a possible local maximum is
similar to a local maximum of $t_{cr}$ in the RSG phase observed in our
system [see Figs.~\ref{fig06}(a) and (b)].  Here we discuss the $T$
dependence of the peak height $S_{max}$.  The peak height $S_{max}$ ($H =
0.5$ Oe and $t_{w} = 1.0 \times 10^{3}$ sec) exhibits a peak at 13 K just
below $T_{RSG}$, having a local minimum at 25 K, and tends to increase with
further increasing $T$.  Although there have been no data on $S(t)$ above
30 K, it is assumed that $S_{max}$ shows a local maximum between 30 K and
$T_{c}$, since $S_{max}$ should reduce to zero above $T_{c}$.  These two
peaks of $S_{max}$ vs $T$ are similar to two local maxima around $T_{RSG}$
and between $T_{RSG}$ and $T_{c}$ in our system.  Next we discuss the $H$
dependence of the peak height $S_{max}$.  The peak height $S_{max}$ ($t_{w}
= 1.0 \times 10^{3}$ sec) decreases with increasing $H$ at both $T = 16$
and 30 K. The drastic decrease of $S_{max}$ at $T = 30$ K occurs even at $H
= 0.2$ Oe, while the decrease of $S_{max}$ at 16 K with $H$ is much weaker
at low $H$ below 1 Oe.  The $H$-dependence of $S_{max}$ at 16 and 30 K is
similar to that at $T = 3.3$ and 7.0 K for our system (see
Fig.~\ref{fig11}(a)).

The second case is the results of $S(t)$ for Fe$_{0.70}$Al$_{0.30}$
($T_{RSG} = 92$ K and $T_{c} = 400$ K).  Motoya et al.\cite{Motoya2003}
have reported the $t$ dependence of $S(t)$ at various $T$ only in the RSG
phase, where $H$ = 10 Oe and $t_{w} = 3.6 \times 10^{3}$ sec.  The
relaxation rate $S(t)$ shows a peak at $t_{cr}$.  The time $t_{cr}$
decreases with increasing $T$ below $T_{RSG}$, while the peak height
$S_{max}$ exhibits a peak at 50 K well below $T_{RSG}$.  These results are
in good agreement with our results: see Figs.~\ref{fig06}(a) and (b) for
$t_{cr}(T)$ and Fig.~\ref{fig07} for $S_{max}(T)$ having a local maximum in
$S_{max}$ at 3.2 K below $T_{RSG}$.  Motoya et al.\cite{Motoya2003} have
also measured the $t$ dependence of $S(t)$ at 10 K at various $H$, where $T
= 10$ K and $t_{w} = 3.6 \times 10^{3}$ sec.  Both $t_{cr}$ and $S_{max}$ decrease with
increasing $H$.  This result is in good agreement with our results: see
Fig.~\ref{fig10}(a) for $t_{cr}(H)$ and Fig.~\ref{fig11}(a) for
$S_{max}(H)$.

\subsection{\label{disC}Scaling of $\chi_{ZFC}(t)$ and 
$\chi^{\prime\prime}(\omega,t)$}
As is described in Sec.~\ref{backA}, the $t$ dependence of
$\chi_{ZFC}(t)$ and $\chi^{\prime\prime}(\omega,t)$ arises from that
of the spin auto-correlation function $C(t_{w};t+t_{w})$ and $C(\Delta
t_{\omega};t+\Delta t_{\omega})$, respectively, where $\Delta t_{\omega} =
2\pi/\omega$.  Here we assume that $C(t_{w};t+t_{w})$ is defined by either
Eqs.(\ref{eq05}) or (\ref{eq06}).  The function $C_{ag}(t_{w};t+t_{w})$
is approximated by a scaling function $F(t/t_{w})$, where $F(x)$ has a
power-law form $x^{-b}$ for $x \gg 1$.  In the present work, $t_{w}$ takes
various values as $\Delta t_{\omega} =1/f = 10^{-3}$ - 100 sec ($0.01 \leq
f\leq 1000$ Hz) for $\chi^{\prime\prime}(\omega,t)$ and as a wait time
$t_{w} = (0.2 - 3.0) \times 10^{4}$ sec for $\chi_{ZFC}(t)$.  The
function $C_{eq}(t)$ is independent of $t_{w}$ and is expressed by a
power-law form ($\approx t^{-\alpha}$) given by Eq.(\ref{eq07}).  By the
appropriate choice of $t$ and $t_{w}$ (or $\Delta t_{\omega}$), it is
experimentally possible to separate the quasi-equilibrium part $C_{eq}(t)$
and the aging part $C_{ag}(t_{w};t+t_{w})$.  Our experimental results are
as follows.  (i) The absorption $\chi^{\prime\prime}(\omega,t)$ mainly
comes from the quasi-equilibrium contribution $C_{eq}(t)$.  This is also
supported by the fact that no $\omega t$-scaling is observed for
$\chi^{\prime\prime}(\omega,t)$.  The exponent $\alpha$ is positive and
very small; typically $\alpha = 0.07$ at $T = T_{RSG}$ = 3.5 K in the RSG
phase (see Table \ref{table01}).  This value of $\alpha$ is considered to
coincide with the exponent $m$ for the stretched exponential relaxation. 
(ii) The susceptibility $\chi_{ZFC}(t)$ for $t \approx t_{w}$ and $t
> t_{w}$ mainly comes from the aging contribution.  As shown in
Fig.~\ref{fig08}, the exponent $n$ depends on $T$ and is 0.78 at $T \approx
T_{RSG}$.

Ozeki and Ito\cite{Ozeki2001} have studied the nonequilibrium relaxation of
the $\pm J$ Ising model in three dimensions.  They have shown that
$C(t_{w};t+t_{w})$ obeys the power-law form: $t^{-\lambda q}$ for $t\ll
t_{w}$ and $t^{-\lambda_{ne}}$ and for $t \gg t_{w}$, where $\lambda_{q} =
0.070(5)$ and $\lambda_{ne} = 0.175(5)$ at $T = 0.92T_{SG}$.  The value of
$\lambda_{q}$ is consistent with that obtained by
Ogielski:\cite{Ogielski1985} $\lambda_{q} = 0.065$ at $T =
T_{SG}$.  The exponent $\lambda_{ne}$ characterizes the nonequilibrium
relaxation.  The exponent $\lambda_{q}$ at $T_{SG}$ is described by
$\lambda_{q}(T_{SG}) =\beta/x$, where $x$ is the dynamic critical exponent
and $\beta$ is the exponent of the SG order parameter.  In the previous
paper\cite{Suzuki2004} we have reported the values of $\beta$ and $x$ as
$\beta = 0.57$ and $x = 8.5$, leading to $\lambda_{q} = 0.067$.  The
agreement between the theory and experiment is very good.  It is predicted
that the $t$ dependence of $\chi^{\prime\prime}(\omega,t)$ for $\omega 
t \gg 1$ is dominated by that of the aging part
$\chi_{ag}^{\prime\prime}(\omega,t)$: $\chi_{ag}^{\prime\prime}(\omega,t)
\approx (\omega t)^{-b}$ with $b = \lambda_{ne}$.  Experimentally it has
been confirmed that $\chi^{\prime\prime}(\omega,t)$ in the SG phase obeys
the $\omega t$-scaling law with $b = 0.14 \pm 0.03$ for
Fe$_{0.5}$Mn$_{0.5}$TiO$_{3}$ ($T_{SG} = 20.7$ K)\cite{Dupuis2001} and $b =
0.255 \pm 0.005$ for Cu$_{0.5}$Co$_{0.5}$Cl$_{2}$-FeCl$_{3}$
GBIC.\cite{Suzuki2003b}

\subsection{\label{disD}Validity of the stretched exponential relaxation}
We discuss the validity of the stretched exponential relaxation.  We 
show that $\chi_{ZFC}(t)$ can be well described by the stretched
exponential relaxation with $m = 0$.  However, we find that the
least squares fit of the data of $\chi_{ZFC}$ vs $t$ to the stretched
exponential relaxation with $m \neq 0$ does not work well in determining
the values of $n$ and $m$.  The value of $n$ is very sensitive to the small
change in $m$.  In Sec.~\ref{backB} we assume that $\chi_{ZFC}(t)$ is given
by Eq.(\ref{eq16}).  Then $S(t)$ has a peak at $x_{cr} = t_{cr}/\tau$
described by Eq.(\ref{eq20}).  As shown in the contour plot of $x_{cr}$ in
the ($n$,$m$) plane of Fig.~\ref{fig01}(a), the value of $x_{cr}$ strongly
depends on the values of $n$ and $m$.  In Fig.~\ref{fig01}(b) we show the
plot of $x_{cr}$ as a function of $m$ for a fixed $n$.  The value of
$x_{cr}$ takes 1 at $m = 0$ and drastically decreases with increasing $m$. 
Experimentally the value of $x_{cr}$ is estimated from the ratio of
$t_{cr}$ to $\tau$, where $t_{cr}$ is the time at which $S(t)$ takes a peak
and $\tau$ is from the least squares fit of the data of $\chi_{ZFC}$ vs $t$
to the stretched exponential relaxation with $m = 0$.  The ratio $x_{cr}$
provides a good measure to determine whether the stretched exponential
relaxation is valid for our system.  Experimentally we find that the ratio
$x_{cr}$ is dependent on $t_{w}$ and $T$ ($3 \leq T \leq 8$ K): $x_{cr} =
0.93 \pm 0.29$ for $t_{w} = 1.5 \times 10^{4}$ sec and $x_{cr} = 1.27 \pm
0.15$ for $t_{w} = 3.0 \times 10^{4}$ sec.  In our simple model, it follows
that $m$ becomes negative when $x_{cr}>1$.  Furthermore, in 
Sec.~\ref{disC}, we show
that $n = 0.78$ and $m = 0.07$ at $T_{RSG}$.  However, these values do not
satisfy the inequality ($4m+n<1$).  This inequality is required for the
peak of $S(t)$ to appear in the case of the stretched exponential
relaxation.  When $4m+n = 1$, $x_{cr} = 2^{-2/1-n}$, which is independent
of $m$.  The value of $x_{cr}$ becomes zero as $n$ tends to unity.  In our
system, $n$ = 0.73 - 0.81.  Using the inequality [$m < (1-n)/4$], the upper
limit of $m$ can be estimated as 0.0475 for $n = 0.81$ (or $1-n = 0.19$). 
Here we note that the exponent $b$ for $\chi_{ag}^{\prime\prime}(\omega,t)$
[$\approx (\omega t)^{-b}$] is on the same order as the value of $(1-n-m)$
or $(1-n)$.  In general the relationship between $n$ and $b$
may be derived from the relaxation rate $S(t)$ as a function $t/t_{w}$,
where $t \approx t_{w}$ is replaced by $2\pi/\omega$.

\section{\label{conc}CONCLUSION}
Cu$_{0.8}$Co$_{0.2}$Cl$_{2}$-FeCl$_{3}$ GBIC undergoes successive
transitions at the transition temperatures $T_{c}$ ($\approx 9.7$ K) and
$T_{RSG}$ ($\approx 3.5$ K).  The FM phase of our system is characterized
by the aging phenomena and nonlinear magnetic susceptibility.  The
relaxation rate $S(t)$ exhibits a characteristic peak at $t_{cr}$ close to
a wait time $t_{w}$ below $T_{c}$, indicating the occurrence of aging
phenomena in both the RSG and FM phases.  The aging behavior in the FM
phase is fragile against a weak magnetic-field perturbation.  In the FM
phase there occurs a partial rejuvenation effect in $S(t)$ under the $T$-shift
perturbation.  The time ($t$) dependence of $\chi_{ZFC}(t)$ around $t$
$\approx t_{cr}$ is well approximated by a stretched exponential
relaxation.  The relaxation time $\tau$ ($\approx t_{cr}$) exhibits a local
maximum around 5 K, reflecting a chaotic nature of the FM phase.  It
drastically increases with decreasing temperature below $T_{RSG}$, as is
usually seen in the SG phase of SG systems.

\begin{acknowledgments}
We would like to thank H. Suematsu for providing us with single
crystal kish graphite, and T. Shima and B. Olson for their
assistance in sample preparation and x-ray characterization. Early
work, in particular for the sample preparation, was supported by
NSF DMR 9201656.
\end{acknowledgments}

\end{document}